\documentclass{tlp}
\usepackage{eepic}
\usepackage{epic}
\usepackage{latexsym}

\long\def\ignore#1{}

\newcommand{\ra}{\rightarrow}
\newcommand{\lra}{\longrightarrow}
\newcommand{\sep}{\;\vert\;}
\newcommand{\app}{{\ }}
\newcommand{\lambdax}[1]{\lambda #1 \, }
\newcommand{\allx}[1]{\forall #1 \,}
\newcommand{\somex}[1]{\exists #1 \,}

\newcommand{\subfor}[3]{#3[#2 := #1]}
\newcommand{\tup}[1]{{\langle#1\rangle}}

\newcommand{\dum}[1]{@ #1}
\newcommand{\lenv}{{\lbrack\!\lbrack}}
\newcommand{\renv}{{\rbrack\!\rbrack}}
\newcommand{\env}[1]{{\lenv #1 \renv}}

\newcommand{\pderivation}{{$\cal P$}\kern -.1em\hbox{-derivation}}
\newcommand{\pderivable}{{$\cal P$}\kern -.1em\hbox{-derivable}}
\newcommand{\simpl}{{\it SIMPL}}
\newcommand{\match}{{\it MATCH}}
\newcommand{\failed}{{\bf F}}
\newcommand{\Pscr}{{\cal P}}
\newcommand{\Gscr}{{\cal G}}
\newcommand{\Dscr}{{\cal D}}

\newcommand{\ie}{{\em i.e.}}
\newcommand{\eg}{{\em e.g.}}

\newenvironment{example}{\refstepcounter{exno}\medskip \noindent {\bf
Example \theexno.\ }}{\hspace*{\fill}$\Box$\medskip}

\newenvironment{defn}{\refstepcounter{defno}\medskip \noindent {\bf
Definition \thedefno.\ }}{\hspace*{\fill}$\Box$\medskip}

\newcommand{\etc}{{\em etc}}

\begin{document}

\title[A Treatment of Higher-Order Features in Logic Programming]
{A Treatment of Higher-Order\\ Features in Logic Programming}

\author[G. Nadathur]
{GOPALAN NADATHUR\thanks{This work has been partially supported by the
    National Science Foundation under Grants CCR-9803849 and
    CCR-0096322. Support was also received from the Digital Technology
    Center and the Department of Computer Science and Engineering at
    the University of Minnesota. The final version of the paper was
    prepared during a sabbatical visit to the Protheo
    group at LORIA and INRIA, Nancy.}\\
Digital Technology Center and Department of Computer Science and Engineering\\
University of Minnesota\\
4-192 EE/CS Building, 200 Union Street S.E.\\
Minneapolis, MN 55455\\
E-mail: gopalan@cs.umn.edu
}

\maketitle

\begin{abstract}

The logic programming paradigm provides the basis for a new
intensional view of higher-order notions. This view is realized
primarily by employing the terms of a typed lambda calculus as
representational devices and by using a richer form of unification
for probing their structures.  These additions have important
meta-programming applications but they also pose non-trivial
implementation problems. One issue concerns the machine representation
of lambda terms suitable to their intended use: an adequate encoding
must facilitate comparison operations over terms in addition to
supporting the usual reduction computation. Another aspect relates to
the treatment of a unification operation that has a branching
character and that sometimes calls for the delaying of the solution of
unification problems. A final issue concerns the execution of goals
whose structures become apparent only in the course of computation.
These various problems are exposed in this paper and solutions to them
are described. A satisfactory representation for lambda terms is
developed by exploiting the nameless notation of de Bruijn as well as
explicit encodings of substitutions. Special mechanisms are
molded into the structure of traditional Prolog implementations to
support branching in unification and carrying of unification
problems over other computation steps; a premium is placed in this
context on exploiting determinism and on emulating usual first-order
behaviour. An extended compilation model is presented that
treats higher-order unification and also handles dynamically emergent
goals. The ideas described here have been employed in the {\it Teyjus}
implementation of the $\lambda$Prolog language, a fact that is used to
obtain a preliminary assessment of their efficacy.

\end{abstract}


\section{Introduction}\label{sec:intro}

Customary acquaintance with higher-order notions in programming
relates to the imperative or functional programming paradigms. Within
these frameworks, functions are equated with the methods for computing
that are embodied in procedures. Higher-orderness then consists of the
ability to objectify such functions and, thereby, to embed them in
data or pass them as arguments to other functions. Many interesting
applications have been found for such capabilities. However, all of
these are dependent on a uniform {\it extensional} view of functions.
In particular, the only observable aspect of such objects is their
ability to transform given arguments into result values.

Logic programming has the potential for supporting a different and
more sophisticated understanding of higher-order notions
\cite{NM98}. Functions are used within this paradigm as a means for
constructing descriptions of objects. Such descriptions can be
examined by means of unification, an operation that is useful in the
analysis of {\it intensions}. Traditional logic programming languages
manifest a weak exploitation of this capability because they
permit only individual, non-function, objects as values. However, it
is possible to support the probing of function structure in genuinely
higher-order ways by introducing a mechanism such as the terms of a
lambda calculus for encoding function objects and by complementing
this with richer notions of variables and unification.  The usual form
of higher-order programming can be realized simply by using the
ability to represent function valued objects and the extensional
interpretation built into logic programming of one kind of function,
namely, predicates. The richer intensional view of functions offers,
in addition, many possibilities that have not been systematically
supported by any previous programming paradigm. To consider one 
important direction, the ability to use lambda terms as
representational devices lends itself well to an abstract view of
syntax that treats binding notions explicitly \cite{MN87slp,PE88pldi},
leading thereby to many novel metalanguage applications for logic
programming (\eg, see 
\cite{AppFel99,Felty93jar,hannan92lics,Pfe88,HM92mscs}). 

While there is considerable application potential for higher-order
features in logic programming, the addition of such features also
raises significant implementation problems.  One category of problems
arises from the use that is made of the terms of a lambda calculus
essentially as data structures. This is a truly novel role for such
terms in programming, and a representation must be developed for them
that supports their use in this capacity. A satisfactory representation
should permit the examination of term structure and must facilitate
the comparison of terms in a situation where the particular names of
bound variables are unimportant, in addition to efficiently supporting
the usual reduction operation on terms. Another class of problems
relates to the fact that the unification computation on lambda terms,
known as higher-order unification \cite{Huet75}, possesses
characteristics that are distinct from those of the customary
first-order unification. In particular, performing this operation may
involve a branching search and it may also be necessary to temporarily
`suspend' the computation before a unifier is found. 
Suitable dynamic support must be described for both facets. 
A third aspect that needs special consideration is the mixing of
intensional and extensional views of predicate objects. There is a
distinction between examining the structure of an object and using
this structure to determine the invocation of code.  A satisfactory
method must be provided for realizing the switch between these
roles. Finally, any machinery that is designed for supporting the new
features 
must be interwoven into the run-time devices and approaches to
compilation that are commonly used in logic programming
implementation. The proper combination of all these mechanisms in one
system is itself a non-trivial issue.  

We consider all these problems in this paper and we develop
methods for addressing them. For the sake of concreteness, we describe
our new implementation ideas within the
framework of the Warren Abstract Machine (WAM) \cite{War83}, a popular
vehicle for realizing logic programming languages.  One of our
contributions relates to the representation of lambda
terms. We carefully identify the different issues that
become relevant where these terms are used intensionally and we
develop an encoding for them that utilizes mechanisms for eliminating
bound variable names \cite{debruijn72} and for capturing substitutions
in terms \cite{NW98tcs} towards addressing these issues. We also
outline a low-level realization of such an encoding and we discuss
the integration of operations on these terms into an abstract machine
structure.  Another contribution is the development of machinery for
supporting the special needs of higher-order unification.  In this
direction we, first of all, describe an explicit encoding of
unification problems that exploits the manner in which these evolve to
foster sharing in their representation.  We also propose mechanisms
that are suitable for realizing branching in unification through a
depth-first search regimen with the possibility of backtracking.
Finally, recognizing that branching search is, in general,
computationally expensive, we describe a processing structure that
facilitates the application of special deterministic steps and that
delays the consideration of branching until after such steps. Using
this approach it is possible to treat first-order
unification almost exactly as it would be treated in a Prolog
implementation, a facet recognized to be important even in
higher-order programs \cite{MP92}. The final contribution of this
paper relates to compilation. We propose enhancements to the structure
of the WAM and modifications to its instruction set that together
realize a compiled execution of programs in a higher-order
language. We also outline in this context a treatment of the 
transition between intensional and extensional roles of predicates. 

The ideas that we develop in this paper have a special practical
relevance: they are useful to the implementation of the logic
programming language $\lambda$Prolog \cite{NM88}. This language
actually embodies two extensions to a Prolog-like language in addition
to the higher-order features considered here. In one direction, it
makes richer use of logical connectives and quantifiers to introduce
notions of scoping \cite{MNPS91}. In another direction, it includes a
polymorphic typing regimen \cite{NP91}. Both aspects raise new
questions for implementation that we have addressed elsewhere
\cite{KNW92,NJK94}. The machinery that we describe here blends well
with these other mechanisms and all our ideas have, in fact, been
amalgamated in a new implementation of $\lambda$Prolog called {\it
Teyjus} \cite{NM99cade}.

The rest of this paper is structured as follows.  In the next section
we identify a logic programming language that embodies the
higher-order features that are presently of interest and we
characterize computation in this language. 
In Section~\ref{sec:interp}, we refine the description of computation
into one that provides the basis for implementation by outlining the
structure of the higher-order unification operation and using
this to develop an abstract interpreter for the language.
The remainder of this paper concerns a low-level
realization of this interpreter. In Section~\ref{sec:terms}, we
discuss issues relevant to the representation of lambda terms and
distill from this an encoding for them that can be used in an actual
implementation. The following section integrates our term
representation into an overall computational model and proposes new
machinery for the realization of higher-order unification.
Section~\ref{sec:abstmachine} makes explicit an extended abstract
machine structure and considers the compilation of first-order
like unification as well as the treatment of higher-order aspects
relative to  
this machine. Section~\ref{sec:conc} discusses related work and
concludes the paper.

\section{A Higher-Order Language}\label{sec:hohc}

The logical language whose implementation we consider in this paper is
an analogue within Church's Simple Theory of Types
\cite{Church40} of the Horn clause language that underlies
Prolog. Church's 
logic is one that builds on a typed lambda calculus. In the
interpretation we use here, the types are constructed from given sets
of sorts and type constructors, each element of the latter set being
attributed a specific arity. The set of sorts initially contains $o$,
the type of propositions, and others such as $int$, $real$, \etc, with
obvious interpretations. We also assume the availability of at least
the unary list type constructor $list$. Both these sets must be
augmentable dynamically in a programming situation, a fact that we
will utilize implicitly.  The full collection of types is the smallest
set satisfying the following properties: (i)~each sort is a type,
(ii)~if $\alpha_1,\ldots,\alpha_n$ are types and $c$ is a type
constructor of arity $n$, then $(c\app \alpha_1\app\ldots\app
\alpha_n)$ is a type, and (iii)~if $\alpha$ and $\beta$ are types,
then so is $(\alpha 
\ra \beta)$. A function type is one whose top-level structure
has the form $(\alpha \ra \beta)$. All other types are considered to
be atomic. We minimize the use of parentheses by assuming that the
application of a type constructor has highest 
priority and that $\ra$ is right associative. The latter
convention allows any function type to be written in the form
$\alpha_1 \ra 
\ldots \ra \alpha_n \ra \beta$ where $\beta$ is an atomic type. The
target type of such a type is $\beta$ and
$\alpha_1,\ldots,\alpha_n$ are its argument types. This notation
and terminology is extended to atomic types by permitting the argument
types to be an empty sequence. 

Starting from typed collections of constants and variables, the terms
in the language are identified together with their types via the
following rules:
(i)~each constant and variable of type $\alpha$ is a term of type
$\alpha$, (ii)~if $x$ is a variable of type $\alpha$ and $t$ is a term
of type $\beta$, then $(\lambdax {x} t)$ is a term of type $\alpha
\ra \beta$ and is called an abstraction that binds $x$ and has scope
$t$, and (iii)~if $t_1$ and $t_2$ are terms of type $\alpha \ra \beta$
and $\alpha$ respectively, then $(t_1 \app t_2)$ is a term of type
$\beta$ and is called an application of $t_1$ to $t_2$. We
reduce the use of parentheses in writing terms by assuming that
application is left associative and that an abstraction has as its
scope the largest well-formed term to the right of the variable it
binds.

The constants in the language are partitioned into {\it non-logical}
and {\it logical} ones. The former set contains an initial
collection of elements such as those representing the integers and
is augmentable in some manner that we, once again, leave
implicit. The set of logical constants consists of the symbols $\top$
of type $o$ denoting the tautologous proposition, $\neg$ of type $o
\ra o$ denoting negation, $\land$, $\lor$ and $\supset$ of type $o \ra
o \ra o$ denoting conjunction, disjunction and implication,
respectively, and, for each $\alpha$, $\Sigma_\alpha$ and $\Pi_\alpha$
of type $(\alpha \ra o) \ra o$.  The last two `families' of constants
represent generalized existential and universal quantifiers: formulas
usually written as $\somex {x} P$ and $\allx {x} P$ are rendered in
this logic as $\Sigma_\alpha\app
\lambdax {x} P$ and $\Pi_\alpha\app \lambdax {x} P$ for an appropriate
type $\alpha$. We will, in 
fact, use the former as abbreviations of the latter. Although type
subscripts are strictly necessary with $\Sigma$ and $\Pi$, we will
omit these when their identity is obvious or irrelevant to the
discussion at hand. We will also adopt the customary infix notation
for the application of $\land$, $\lor$ and $\supset$ to two
arguments in succession.

We assume the usual notions of free and bound variables and of
subterms of a term. Equality between terms incorporates the rules of
lambda conversion. Let us say that a term $s$ is free for the variable
$x$ in the term $t$ if $x$ does not appear free in $t$ in the  
scope of an abstraction that binds a free variable of $s$. Further,
let $\subfor{s}{x}{t}$ denote the result of replacing all
the free occurrences of $x$ in $t$ by $s$. The lambda conversion rules
that we use are then the following:

\begin{enumerate}

\item ($\alpha$-conversion) Replacing a subterm of the form $\lambdax
{x} t$ in a given term by $\lambdax {y} (\subfor{y}{x}{t})$, provided
$y$ is a variable of the same type as $x$ that is neither free for $x$
in $t$ nor free in $t$. 

\item ($\beta$-conversion) Replacing a subterm of the form $(\lambdax
{x} t_1)\app t_2$ in a given term by $\subfor {t_2} {x} {t_1}$ or
vice versa, provided $t_2$ is free for $x$ in $t_1$. 

\item ($\eta$-conversion) Replacing a subterm of the form $t$ in a
given term by $\lambdax {x} t \app x$ and vice versa, provided $t$
is of type $\alpha \ra \beta$ and $x$ is a variable of type $\alpha$
that is not free in $t$. 

\end{enumerate}

\noindent Two terms are considered to be equal if one can be
obtained from the other by using a sequence of these rules. In
determining such equality, it is often necessary to consider directed
applications of these conversion rules. Of particular importance is an
oriented form of the $\beta$-conversion rule that is made precise as
follows. First, we identify a term of the form $(\lambdax
{x} t_1)\app t_2$ as a {\it $\beta$-redex}; in the sequel, we shall
also refer to such a term more simply as just a redex and shall
call $t_1$ its body and $t_2$ its argument. Now, the condition
permitting the replacement of a subterm 
of this kind as per the $\beta$-conversion rule may
not be satisfied in general, but this can be corrected by using a
sequence of $\alpha$-conversion steps. We call such a sequence
followed by the desired application of the $\beta$-conversion rule a
$\beta$-contraction.

We will need to consider the idea of unifying two lambda terms. The
interest here is in substituting terms of matching types for free
variables so that the two terms become equal. This substitution
operation must be performed with care to 
ensure that free variables in the substituted terms do not get
accidentally bound in the result. A correct characterization of this
operation can, in fact, be provided using equality modulo the lambda
conversion rules. Thus, suppose that, for $1 \leq i \leq n$, $t_i$ and
$x_i$ are a term and a variable of identical type. Then, the set
$\{\langle x_i, t_i\rangle \vert 1 \leq i \leq n \}$ represents a
substitution and the application of this substitution to $t$ is
equal to the term $(\lambdax {x_1} \ldots \lambdax {x_n} t)\app t_1
\app \ldots \app t_n$.

A central part of generalizing Horn clauses to higher-order logic is
describing a suitable notion of atomic formulas. Towards this end, we
first identify the class of {\it positive} terms as those lambda terms
in which the only logical constants that appear are
$\land$, $\lor$ and $\Sigma_\alpha$. Our atomic formulas or atoms are
then all the terms of type $o$ of the form $P\app t_1 \app \ldots \app
t_n$ where $P$ is a (predicate) variable or non-logical constant and,
for $1 \leq i \leq n$, each $t_i$ is a positive term. Such a formula
is said to be {\it rigid} 
just in case $P$, its predicate head, is a non-logical constant and is
said to be {\it flexible} otherwise. We denote arbitrary atoms
by $A$ and rigid ones by $A_r$ below. {\it Goal formulas} or
simply {\it goals} are then the propositional terms that are denoted
by $G$ in the syntax rule 
\[G ::= \top \sep A \sep G\land G \sep G\lor G \sep \somex {x} G.\]
These formulas are higher-order versions of queries or goals in
Prolog; notice, in particular, that the arguments of atomic goals are
lambda terms as opposed to first-order terms and predicate and
function variables  are permitted in goals. A higher-order Horn clause
or program clause is the universal closure of a term of the form $A_r$
or $G \supset A_r$. Program clauses are intended to
be interpreted in a computational setting as (partial) definitions of
procedures and from this perspective the restriction to rigid atoms
is well-motivated: such an interpretation is meaningful only
if the `procedure' has a definite name. 

A multiset of higher-order program clauses constitute a program in our
logic programming language. Computation is engendered by a
goal formula being presented relative to a given program.  Such a goal
formula typically has existential quantifiers at its head and the
programming task is to find instantiations for the quantified
variables that permit the resulting goal to be solved from the
program. 

\begin{example}\label{ex:mapfun}
Let $i$ be a sort representing individuals and let the set of
nonlogical constants contain the following: 
\begin{tabbing}
\qquad\=${\it mapfun}$\quad\=\kill
\>${\it nil}$\>of type $list\app i$\\
\>$::$\> of type $i \ra list\app i \ra list \app i$, and\\
\>${\it mapfun}$\> of type $list \app i\ra (i \ra i) \ra list \app i\ra
o$
\end{tabbing}
\noindent Further, assume that $::$ can be written as an infix, right
associative operator. Then the
clauses
\begin{tabbing}
\qquad\=\kill
\>$\allx {f}({\it mapfun}\app {\it nil}\app f\app {\it nil})$\ and\\
\>$\allx {x} \allx {f} \allx {l_1} \allx {l_2}({\it mapfun}\app l_1\app f\app
l_2) \supset ({\it mapfun}\app (x \app ::\app l_1)\app f\app ((f \app x) \app
:: \app l_2))$
\end{tabbing}
\noindent constitute a program.\footnote{The omission of types with
the quantifiers in these clauses illustrates the convention alluded to
earlier. Here, the type of ${\it mapfun}$ uniquely
determines the types of the quantified variables.} Adopting Prolog's
suggestive manner of writing implications, its convention of making
quantifiers 
implicit by using names beginning with uppercase letters for
quantified variables and its method for depicting each program clause,
this program can be rendered into the following `friendlier' syntax:  
\begin{tabbing}
\qquad\=\kill
\>${\it mapfun} \app {\it nil}\app F\app {\it nil}.$\\
\>${\it mapfun}\app (X :: L1)\app F\app ((F\app X) :: L2)$ {\tt :-}
${\it mapfun}\app L1\app F\app L2.$ 
\end{tabbing}
\noindent Letting $g$ be a constant of type $i \ra i \ra i$ and $a$
and $b$ be constants of type $i$, the
formula $\somex {l} {\it mapfun}\app (a :: b :: {\it nil})\app (\lambdax {x} g
\app a\app x) \app l$ 
constitutes a query. Using Prolog's conventions for making
quantifiers implicit, this query may be rewritten as 
\begin{tabbing}
\qquad\=\kill
\> ${\it mapfun}\app (a :: b :: {\it nil})\app (\lambdax {x} g \app
a\app x) \app L$.
\end{tabbing}
\noindent There is exactly one solution to this query, this being
given by the `answer' substitution $\{\langle L,(g\app a\app a) ::
(g\app a \app b) :: {\it nil}\rangle\}.$ 
Notice that generating this answer substitution requires, amongst
other things, the application of a lambda term to two different
arguments and the subsequent reduction of these terms to normal
form. An alternative query is the following:
\begin{tabbing}
\qquad\=\kill
\> ${\it mapfun}\app (a :: b :: {\it nil})\app F \app ((g\app a \app
a) :: (g \app a \app b) :: {\it nil})$.
\end{tabbing}
\noindent This query also has a unique solution, this being the
substitution $\{\langle F, \lambdax {x} g \app a \app x \rangle \}$.
Computing this answer involves unifying two pairs of terms containing a
function variable, these being $F\app a$ and $g\app a \app a$ on the
one hand and $F\app b$ and $g\app a \app b$ on the other. We discuss
in the next section a process by which such unification
problems may be solved. For the moment, we note that the first of the
pairs of terms has four distinct most general unifiers given by the
following substitutions for $F$: 
\begin{tabbing}
\qquad\=\kill
\>$\{\langle F,\lambdax {x} g\app a\app a \rangle \}$, $\{\langle F,
\lambdax {x} g\app x\app a\rangle \}$, $\{\langle F, \lambdax {x}
g\app x \app x\rangle \}$, and $\{\langle F,\lambdax {x} g\app a\app x
\rangle \}$.
\end{tabbing}
\noindent If the two pairs of terms in question are unified sequentially
and any but the last solution is chosen initially for the first pair,
then it will be necessary to backtrack to find a solution for the
composite problem.
\end{example}

We shall employ the conventions used for depicting formulas
in the above example freely in the rest of the paper. The predicate
${\it mapfun}$ considered in this example relates a function and two lists
just in case the second list is obtained by applying the function to
each element of the first list. The notion of function application is,
however, relatively weak, being given by reduction in a typed lambda
calculus with no interpreted constants. A stronger form of function
application, one that invokes the ability of solving goals in the
underlying language, can be realized by using a predicate version of
${\it mapfun}$ as described in the following example.

\begin{example}\label{ex:mappred}
In addition to the constants and types of Example~\ref{ex:mapfun},
assume that ${\it mappred}$ is a constant of type  
$list\app i \ra (i \ra i \ra o) \ra list\app i \ra o$.
Then, using the Prolog convention of depicting conjunctions
by commas, the following clauses correspond to a program:
\begin{tabbing}
\qquad\=\kill
\>${\it mappred}\app {\it nil}\app P\app {\it nil}$.\\
\>${\it mappred}\app (X :: L1)\app P\app (Y :: L2)$ {\tt :-}
$(P\app X\app Y),({\it mappred}\app L1\app P\app L2).$
\end{tabbing}
\noindent Let ${\it bob}$, ${\it john}$, ${\it mary}$, ${\it sue}$, ${\it dick}$ and ${\it kate}$ be
constants of type $i$ and let ${\it parent}$ be a constant of type $i \ra i
\ra o$. Then the following additional clauses define a `parent'
relationship between different individuals:
\begin{tabbing}
\qquad\=\kill
\> ${\it parent} \app {\it bob}\app {\it john}$.\\
\> ${\it parent} \app {\it john} \app {\it mary}$.\\
\> ${\it parent} \app {\it sue} \app {\it dick}$.\\
\> ${\it parent} \app {\it dick} \app {\it kate}$.
\end{tabbing}
\noindent In this context, the following term constitutes a query:
\begin{tabbing}
\qquad\=\kill
\> ${\it mappred}\app ({\it bob} :: {\it sue} :: {\it nil}) \app {\it parent} \app L$.
\end{tabbing}
\noindent The sole answer to this query is the substitution
$\{\langle L, {\it john} :: {\it dick} :: {\it nil}\rangle \}$. 
In solving this query, two new goals of the form
$({\it parent}\app {\it bob} \app Y1)$ and $({\it parent} \app {\it
  sue} \app Y2)$ will have to be dynamically formed and
solved. Another example of a query is 
\begin{tabbing}
\qquad\=\kill
\> ${\it mappred} \app ({\it bob} :: {\it sue} :: {\it nil}) \app (\lambdax {x} \lambdax {y}
\somex {z} ({\it parent} \app x \app z) \land ({\it parent} \app z \app y))\app L$.
\end{tabbing}
\noindent This goal asks for the grandparents of ${\it bob}$
and ${\it sue}$ and has as its solution the substitution $\{\langle L,
{\it mary} :: {\it kate} :: {\it nil}\rangle \}$. 
\ignore{
} Finding this answer requires two new goals with complex
structures---each with an embedded conjunction and 
existential quantifier---to be constructed at runtime and then solved.
\end{example}

Example~\ref{ex:mappred} motivates the particular structure chosen for
atomic formulas in the definition of our higher-order logic
programming language. Logical constants that appear in the arguments of
predicate expressions can become top-level symbols in a goal
constructed at runtime. These constants must, therefore, be limited
to ones that can legitimately appear in such a position, a requirement
that is achieved by the restriction to positive terms. In a
different direction, in contrasting this example with
Example~\ref{ex:mapfun}, a question that arises is whether or
not the ${\it mappred}$ predicate can be run in `reverse'. For example, is
the query
\begin{tabbing}
\qquad\=\kill
\> ${\it mappred}\app ({\it bob} :: {\it sue} :: {\it nil}) \app P \app ({\it john} :: {\it dick} :: {\it nil})$
\end{tabbing}
\noindent computationally meaningful? It is tempting to decide that
it is and that it has $\{\langle P, {\it parent}\rangle \}$ as an
answer substitution. However, a little thought reveals that there are
too many relations that are 
true of ${\it bob}$ and ${\it john}$ on the one hand and ${\it sue}$
and ${\it dick}$ on the 
other and so this query is, in a sense, an ill-formed one. We note that
the `solution' $\{\langle P,\lambdax {x} \lambdax {y} \top\rangle \}$
actually subsumes all others in a logical sense and, consistent with
the present viewpoint, we may treat this as the {\it only} legitimate
answer to the posed query.

The idea of solving a goal that we have discussed only intuitively
thus far can be made logically precise using the notion of provability
in classical logic \cite{NM90}. Operationally, this sanctions a recipe
for solving a {\it closed} goal from a program $\Pscr$ that is based
on the structure of the goal: 

\begin{enumerate}
\item The goal $\top$ is solved immediately. 

\item The goal $G_1 \land G_2$ is solved by solving both $G_1$ and
  $G_2$. 

\item The goal $G_1 \lor G_2$ is solved by solving one of $G_1$ and
  $G_2$. 

\item The goal $\somex {x} G$ is solved by solving $\subfor{t}{x}{G}$
  for some closed positive term $t$.

\item A rigid atomic goal $A_r$ is solved either (a)~by determining
  that it is equal to a ground instance of a clause in $\cal P$, or
  (b)~by finding a ground instance $G \supset A'_r$ of a clause in
  $\cal P$ such that $A_r$ and $A'_r$ are equal and then solving $G$. 
\end{enumerate}

\noindent In this description, a ground instance of a program clause
is generated by substituting closed positive terms for the universally
quantified variables in the clauses. 

The recipe described above clarifies the operational semantics
of our language, but needs refinement to become the basis for
implementation. In particular, it is necessary to eliminate from it
the oracle used for picking an appropriate instance of an
existentially quantified goal and to embed in it some method for
treating the choices that have to be made concerning the disjunct of a
disjunctive goal that is to be solved and the clause that is to be
used to solve an atomic formula. These kinds of issues have 
actually to be dealt with already in a first-order language. In that
context, existential goals are treated by delaying the choice of
specific instantiations till such time that information is available
for making the `right' choices. Thus, the goal $\somex {x} G$ is
transformed into one of the form $\subfor{X}{x}G$ where $X$ is a new
variable that  may be instantiated in the course  
of computation. Actual instantiations for such variables are
determined at the time of solving atomic goals.  Given the atomic goal
$A$, we look for a clause of the form $\allx {y_1}
\ldots \allx {y_n} A'$ or $\allx {y_1} \ldots \allx {y_n} (G' \supset
A')$ that is such that $A$ unifies with the formula that results from
$A'$ by replacing the universally quantified variables with new
variables.  If a clause of this kind is found, then, depending on its
form, either the atomic goal succeeds immediately or the next task
becomes that of solving the resulting instance of $G'$.  With regard
to nondeterminism, the usual solution is to make choices in a
predetermined manner and to reconsider these in case of
subsequent failure. Now, the treatment of the logical
connectives, the sequencing through program clauses and much of the
unification computation can, in fact, be compiled and this
is what is actually done within machine models such as the WAM.

The ideas discussed above have obvious applicability in the
implementation of our language as well. However, their precise
deployment must take into account the higher-order nature of this
language. A detailed exposition of the new problems posed by this
aspect and an integration of their treatment into the basic framework
described above is the subject of the rest of this paper.

Before concluding this section, we comment briefly on the rigidity of
the typing regimen used in our language. The predicates 
${\it mapfun}$ and ${\it mappred}$ as we have defined them here are,
for instance, 
restricted to apply to lists of individuals and 
cannot be used with lists of integers, lists of lists or lists of
function objects. This inflexibility can be alleviated by injecting a
form of polymorphism through the use of type variables.  Thus, with an
appropriate change to the underlying typing scheme, ${\it mapfun}$ may have
been defined to be of type $list\app A \ra (A \ra B) \ra list\app B
\ra o$, where $A$ and $B$ can be instantiated by arbitrary types. A
polymorphism of this kind is, in reality, supported by
$\lambda$Prolog. However, we elide this polymorphism here
because it poses additional implementation problems
that we presently do not wish to consider. For the interested
reader, these problems are discussed for a first-order language in
\cite{KNW92}. The solutions provided therein are entirely
compatible with the implementation methods we develop here for our
simply typed language.  

\section{An Abstract Interpreter}\label{sec:interp}

The desired refinement of our recipe for solving goal formulas
requires an understanding of higher-order unification problems and of
a procedure for their 
solution. Problems of this kind are defined by finite multisets of
pairs of terms in which the two terms in each pair have identical
types. We will refer to a collection of this sort as a {\it
disagreement set}. A solution to, or a unifier for, such a problem
is a substitution whose application to the terms in each pair makes
them equal. Higher-order unification problems are, in the general
case, undecidable ones and also do not admit of finite sets of most
general unifiers. There is, nevertheless, a systematic way to check for
unifiability and to enumerate non-redundant sets of preunifiers in
the process. This method, that is due to Huet \cite{Huet75}, has been
used in several programming systems and has demonstrated a practical
usefulness to higher-order unification despite the theoretical
characteristics of the problem.

Huet's method relies on what is known as a {\it head normal form}
for a term. A term is in this form if it has the structure
$\lambdax {x_1} \ldots \lambdax {x_n} (A\app t_1\app \ldots\app t_m)$
where $A$ is either a constant or a variable. Given such a
term, $A$ 
is called its {\it head}, the abstractions at the front of the term
are collectively called its {\it binder}, $t_1, \ldots t_m$ are called
its arguments, $(A\app t_1\app \ldots\app t_m)$ is called its body and
the term is said to be {\it rigid} if $A$ is a constant or an element
of $\{x_1,\ldots,x_n\}$, and {\it flexible} otherwise. Every
term in  our 
typed language can be transformed into such a form modulo the lambda
conversion rules \cite{Andrews71}. Moreover, the results of applying a 
substitution to  a term and to any one of its head normal forms
are equal under these rules. Thus, we may restrict our attention to
terms in such a form as we henceforth do. 

The unification procedure consists of the repetitive use of two
phases for transforming a given disagreement set into a form for which
it can be decided no unifiers exist or for which unifiability is
evident.  The first of these phases is akin to the term simplification
that is an intrinsic part of first-order unification. Consider two
head normal forms that are of the same type. The binders of
these terms may be distinct both in the choice of variable names and in
length at the outset, but these can be arranged to be identical
through the use of the $\alpha$- and $\eta$- conversion rules. We may
therefore assume that the terms in question are, in fact, of the form
$\lambdax {x_1} \ldots \lambdax {x_n} (A_1\app s_1 \app
\ldots \app s_i)$ and $\lambdax {x_1}\app \ldots\app \lambdax {x_n}
(A_2\app r_1 \app\ldots \app r_j)$ respectively. Now, if both terms
are rigid, it can be seen that they are unifiable only if $A_1$ and 
$A_2$ are identical and, in this case, they have the same unifiers as
the set 
\begin{tabbing}
\qquad\=\kill
\>$\{\langle \lambdax {x_1} \ldots \lambdax {x_n} s_1,
\lambdax {x_1} \ldots \lambdax {x_n} r_1 \rangle, \ldots,\langle
\lambdax {x_1} \ldots \lambdax {x_n} s_i, \lambdax {x_1} \ldots
\lambdax {x_n} r_i \rangle \}$;
\end{tabbing}
\noindent note that the identity of types ensure that $i = j$ if $A_1
= A_2$. Thus, given an arbitrary disagreement set, this observation
can be used either to conclude that it has no unifiers or to reduce it
to another disagreement set with the same unifiers and in which each
pair has at least one flexible term. We assume below that this kind of  
simplification is carried out by a function called \simpl\ that
returns a distinguished value \failed\ in the case that it detects the
impossibility of unification.

One of the possibilities for the value returned by \simpl\ is that it
is a disagreement set that has only `flexible-flexible' pairs.
A set of this kind is known to be unifiable but, in the case that it 
is non-empty, a complete search for its unifiers can be unconstrained
\cite{Huet75}. The best strategy for these sets is 
therefore to treat them as constraints on any further processing or,
if computation is at an end, to present them as such on computed
answers.  

The second phase in unification becomes relevant when \simpl\ returns
a set that has at least one `flexible-rigid' pair. A substitution may be
posited for reducing the difference between the terms in the pair in
this case. Two kinds of elementary substitutions completely cover all
the possible ways of doing this. In
particular, let $t_1$ 
be the flexible term with $F$ as its head and let $t_2$ be the rigid
term with $c$ as its head.  Further, let the types of $F$ and $c$ be
$\alpha_1 \ra \cdots \ra \alpha_k
\ra \beta$ and $\gamma_1 \ra \cdots \ra \gamma_j \ra \beta$
respectively, where $\beta$ is an atomic type. Then
\begin{enumerate}
\item the {\it imitation substitution} is defined only when $c$ is a
constant and is
\begin{tabbing}
\qquad\=\kill
\>$\{ \langle F, \lambdax {w_1} \ldots \lambdax {w_k} (c\app (H_1\app
w_1 \app \ldots \app w_k)\app \ldots \app (H_j\app w_1 \app \ldots
\app w_k)) \rangle \}$,
\end{tabbing}
\noindent assuming that $H_1,\ldots,H_j$ are new free
variables of appropriate types, and 

\item for $1 \leq i \leq k$, the {\it $i^{th}$ projection
substitution} is defined only when $\alpha_i$ is of the form $\beta_1
\ra \cdots \ra \beta_l \ra \beta$ and is
\begin{tabbing}
\qquad\=\kill
\>$\{ \langle F, \lambdax {w_1} \ldots \lambdax {w_k}
(w_i\app (H_1\app w_1 \app \ldots \app w_k)\app \ldots \app (H_l\app 
w_1 \app \ldots \app w_k)) \rangle \}$,
\end{tabbing}
\noindent assuming $H_1,\ldots,H_l$ are new free variables of appropriate
types. 
\end{enumerate}
Notice that these substitutions are determined entirely by the
heads of the flexible and rigid terms in question and they are finite
in number.

The iterative use of the two described phases in unification naturally
involves a search whose structure can be visualized through a {\it
  matching tree} \cite{Huet75}. Figure~\ref{fig:match}
presents such a tree for the unification problem $\{\langle F\app a,
g\app a \app a \rangle \}$ encountered in
Example~\ref{ex:mapfun}.  The arcs in this tree are labelled with the
relevant imitation and projection substitutions and the nodes
represent the result of transforming the set on the prior node by
first applying the substitution on the incoming arc and then carrying
out the simplification embodied in \simpl. The leaves of a matching
tree are labelled either with \failed\ or with a multiset of
flexible-flexible pairs. A solution to the original unification
problem can be obtained by composing the substitutions on the path to
the latter kind of leaf with a unifier for that leaf. In the example
presented, observing that an empty disagreement set has the empty
substitution as its most general unifier, these solutions involve
substituting $\lambdax {x} g\app a\app a$, $\lambdax {x} g \app a\app
x$, $\lambdax {x} g \app x \app a$  and $\lambdax {x}g \app x \app x$
for $F$. A matching tree is exhaustive in that the unifiers
of the leaves of a completely expanded tree can be used in this
fashion to produce all the unifiers of the original set. However,
such a tree can, in principle, include nonterminating branches and
can also have an infinite number of `success' nodes.

\begin{figure}
\begin{center}

\setlength{\unitlength}{0.00070833in}
\begingroup\makeatletter\ifx\SetFigFont\undefined%
\gdef\SetFigFont#1#2#3#4#5{%
  \reset@font\fontsize{#1}{#2pt}%
  \fontfamily{#3}\fontseries{#4}\fontshape{#5}%
  \selectfont}%
\fi\endgroup%
{\renewcommand{\dashlinestretch}{30}
\begin{picture}(6849,3363)(0,-10)
\path(4875,2979)(4050,2454)
\path(4135.133,2543.735)(4050.000,2454.000)(4167.346,2493.115)
\path(5175,2979)(6150,2454)
\path(6030.120,2484.478)(6150.000,2454.000)(6058.566,2537.306)
\path(3450,2004)(2250,1404)
\path(2343.915,1484.498)(2250.000,1404.000)(2370.748,1430.833)
\path(3675,2004)(4800,1404)
\path(4680.000,1434.000)(4800.000,1404.000)(4708.235,1486.941)
\path(1800,879)(2775,354)
\path(2655.120,384.478)(2775.000,354.000)(2683.566,437.306)
\path(5100,879)(4200,354)
\path(4288.537,440.378)(4200.000,354.000)(4318.770,388.551)
\path(5400,879)(6300,354)
\path(6181.230,388.551)(6300.000,354.000)(6211.463,440.378)
\drawline(5175,2979)(5175,2979)
\drawline(5175,2979)(5175,2979)
\path(1500,879)(450,354)
\path(543.915,434.498)(450.000,354.000)(570.748,380.833)
\put(4425,3204){\makebox(0,0)[lb]{\smash{{{\SetFigFont{10}{12.0}{\rmdefault}{\mddefault}{\updefault}$\{\langle
F\app a,g\app a\app a\rangle\}$}}}}}
\put(2850,2154){\makebox(0,0)[lb]{\smash{{{\SetFigFont{10}{12.0}{\rmdefault}{\mddefault}{\updefault}$\{\langle
H_1\app a, a\rangle,\langle H_2 \app a,a\rangle\}$}}}}}
\put(1275,1104){\makebox(0,0)[lb]{\smash{{{\SetFigFont{10}{12.0}{\rmdefault}{\mddefault}{\updefault}$\{\langle
H_2\app a, a\rangle \}$}}}}}
\put(4800,1104){\makebox(0,0)[lb]{\smash{{{\SetFigFont{10}{12.0}{\rmdefault}{\mddefault}{\updefault}$\{\langle
H_2\app a,a\rangle\}$}}}}}
\put(2250,729){\makebox(0,0)[lb]{\smash{{{\SetFigFont{10}{12.0}{\rmdefault}{\mddefault}{\updefault}$\{\langle
H_2,\lambdax{x}x\rangle\}$}}}}}
\put(2700,54){\makebox(0,0)[lb]{\smash{{{\SetFigFont{10}{12.0}{\rmdefault}{\mddefault}{\updefault}$\{\}$}}}}}
\put(300,54){\makebox(0,0)[lb]{\smash{{{\SetFigFont{10}{12.0}{\rmdefault}{\mddefault}{\updefault}$\{\}$}}}}}
\put(4050,54){\makebox(0,0)[lb]{\smash{{{\SetFigFont{10}{12.0}{\rmdefault}{\mddefault}{\updefault}$\{\}$}}}}}
\put(6225,54){\makebox(0,0)[lb]{\smash{{{\SetFigFont{10}{12.0}{\rmdefault}{\mddefault}{\updefault}$\{\}$}}}}}
\put(5850,729){\makebox(0,0)[lb]{\smash{{{\SetFigFont{10}{12.0}{\rmdefault}{\mddefault}{\updefault}$\{\langle
H_2,\lambdax{x}x\rangle\}$}}}}}
\put(6225,2154){\makebox(0,0)[lb]{\smash{{{\SetFigFont{10}{12.0}{\rmdefault}{\mddefault}{\updefault}\failed}}}}}
\put(3600,729){\makebox(0,0)[lb]{\smash{{{\SetFigFont{10}{12.0}{\rmdefault}{\mddefault}{\updefault}$\{\langle
H_2,\lambdax{x}a\rangle\}$}}}}}
\put(0,729){\makebox(0,0)[lb]{\smash{{{\SetFigFont{10}{12.0}{\rmdefault}{\mddefault}{\updefault}$\{\langle
H_2,\lambdax{x}a\rangle\}$}}}}}
\put(5700,2829){\makebox(0,0)[lb]{\smash{{{\SetFigFont{10}{12.0}{\rmdefault}{\mddefault}{\updefault}$\{\langle
F,\lambdax{x}x\rangle\}$}}}}}
\put(2175,2829){\makebox(0,0)[lb]{\smash{{{\SetFigFont{10}{12.0}{\rmdefault}{\mddefault}{\updefault}$\{\langle
F,\lambdax{x}g\app (H_1\app x)\app (H_2\app x)\rangle\}$}}}}}
\put(1650,1779){\makebox(0,0)[lb]{\smash{{{\SetFigFont{10}{12.0}{\rmdefault}{\mddefault}{\updefault}$\{\langle
H_1,\lambdax{x}a\rangle\}$}}}}}
\put(4425,1779){\makebox(0,0)[lb]{\smash{{{\SetFigFont{10}{12.0}{\rmdefault}{\mddefault}{\updefault}$\{\langle
H_1,\lambdax{x}x\rangle\}$}}}}}
\end{picture}
}

\end{center}
\caption{A matching tree for $\{\langle (F\app a),(g\app a \app a)
\rangle \}$}
\label{fig:match}
\end{figure}

Our refinement to the earlier model for solving goal formulas consists
of viewing a state in the process as a composite of a collection of
goals and a disagreement set, the latter component arising from the
attempt to solve atomic goals. Progress through this state space is
made by simplification steps applied either to the goals or to the
disagreement set. In any given case, these steps must be relativized
to a particular program ${\cal P}$. The notion of a \pderivation\
\cite{NM90} that generalizes  SLD derivations described in
\cite{AvE82} for first-order Horn clause logic makes this idea
precise.  

\begin{defn} Let $\Pscr$ be a program, let $\Gscr$ be a symbol for
  multisets of goal formulas that we refer to also as {\it goal sets},
  and let $\theta$ be a symbol for substitutions.
 Further, let $\Dscr$ be a symbol for a disagreement set
or the special value \failed. Finally, let \match\ be a 
function on flexible-rigid disagreement pairs that produces the set of
imitation and projection substitutions for any given pair. 
Then a tuple $\tup{{\Gscr} _2,{\Dscr} _2,\theta _2}$ is said to be 
\pderivable\ from a tuple  
$\tup{{\Gscr} _1,{\Dscr} _1,\theta _1}$ in which $\Dscr_1 \ne \failed$
if it is obtainable from the latter by one of the following
steps:\footnote{We intend $\cup$ and $-$ to be interpreted as multiset 
  operations in these clauses.}

\begin{enumerate}

\item {\it Goal simplification step}: $\theta_2 = \emptyset$,
$\Dscr_2 = \Dscr_1$, and for some $G \in \Gscr_1$ it is the case that

\begin{enumerate}

\item $G$ is $\top$ and $\Gscr_2 = \Gscr_1 - \{G\}$, or

\item $G$ is $G_1 \land G_2$ and $\Gscr_2 = (\Gscr_1 -
\{G\}) \cup \{G_1,G_2\}$, or

\item $G$ is $G_1 \lor G_2$ and, for $i = 1$ or $i = 2$,
$\Gscr_2 = (\Gscr_1 - \{G\}) \cup \{G_i\}$, or

\item $G$ is $\Sigma\app P$ and $\Gscr_2 = (\Gscr_1 - \{G\})
\cup \{P\app Y\}$ where $Y$ is a new variable.

\end{enumerate}

\item {\it Backchaining step:} $\theta_2 = \emptyset$ and, for
some rigid atom $G \in \Gscr_1$ either

\begin{enumerate}

\item $A$ is an atom obtained by instantiating the universal
quantifiers in a clause in $\cal P$ with new variables and $\Gscr_2
= \Gscr_1 -\{G\}$ and $\Dscr_2 =\simpl(\Dscr_1 \cup \{\tup{G,A}\})$, or

\item $G' \supset A$ is obtained by instantiating the universal
quantifiers in a clause in $\cal P$ with new variables and
$\Gscr_2 = (\Gscr_1 -\{G\}) \cup \{G'\}$, and $\Dscr_2 =\simpl(\Dscr_1
\cup \{\tup{G,A}\})$. 

\end{enumerate}

\item {\it Flexible goal solution step:} $G \in \Gscr_1$ is an atomic
goal formula that has the (free) variable $Y$ of type $\alpha_1 \ra
\cdots \ra \alpha_n \ra o$ as its head, and $\theta_2 = \{\langle Y,
\lambdax   {x_1} \ldots \lambdax {x_n}  \top \rangle\}$, $\Gscr_2 =
\theta_2(\Gscr_1 - \{G\})$ and $\Dscr_2 = \simpl(\theta_2(\Dscr_1))$;
the application of a substitution to a goal set and a disagreement set
here and below corresponds to its application to the component terms
of these sets.

\item {\it Unification step:} For some flexible-rigid pair
$\chi \in \Dscr_1$, either $\match(\chi) = \emptyset$ and $\Dscr_2 =
\failed$, or $\theta_2 \in \match(\chi)$ and $\Gscr_2 =
\theta_2(\Gscr_1)$ and $\Dscr_2 = \simpl(\theta_2( \Dscr_1))$.

\end{enumerate}
A sequence of the form $\tup{\Gscr_i,\Dscr_i,\theta_i}_{1 \leq i \leq
n}$ is a \pderivation\ sequence for a goal formula $G$ if 
$\Gscr_1 = \{G\}$, $\Dscr_1 = \emptyset$ and $\theta_1 =
\emptyset$, and for $1 \leq j < n$,
$\tup{\Gscr_{j+1},\Dscr_{j+1},\theta_{j+1}}$ 
is \pderivable\ from $\tup{\Gscr_{j},\Dscr_{j},\theta_{j}}$. Such a
sequence terminates in failure if $\Dscr_n = 
\failed$ and with success if $\Gscr_n = \emptyset$ and $\Dscr_n$ is
either empty or contains only flexible-flexible pairs. In the latter
case, we say that the sequence is a \pderivation\ of $G$. Such a
sequence embodies in it a solution to the query $G$ in the context of
the program $\Pscr$ and the {\it answer substitution} corresponding to
it is obtained by composing
$\theta_n \circ \cdots \circ \theta_1$ with any unifier for $\Dscr_n$
and restricting the resulting substitution to the free variables of
$G$.
\end{defn}

An abstract interpreter for our language may be thought of as a
procedure that, given a program $\Pscr$, attempts to construct a
\pderivation\ for goal formulas. Such an interpreter would function by
trying to extend an existing \pderivation\ and will typically be faced
with alternatives in this process. This interpreter can without loss
of completeness choose to use a unification step whenever one is
applicable. The only choices that are critical are, in fact, those of
the disjunct to use when simplifying a disjunctive goal, the clause to
use in a backchaining step, the substitution to use in a unification
step and the point at which to solve a flexible goal. We assume a
depth-first approach with the possibility of backtracking in the
treatment of the first three aspects. The first two kinds of choices
are present in a first-order language as well and similar methods can
be used for treating them here. In particular, we use a left-to-right
processing order in the treatment of disjunctive goals and we base the
selection of  clauses in a backchaining step on an ordering on the
multiset determined by their presentation sequence. Moreover, the
cases with a potential of success can be considerably narrowed down
by techniques such as indexing on predicate names and the structure of
arguments, a fact that we utilize in
Section~\ref{sec:abstmachine}. The treatment of choices in the
unification step and the bookkeeping mechanisms for realizing
backtracking relative to these is a matter we discuss in a later
section. Finally, we assume an initial ordering on goals that we
maintain through an ordered, `in-place' insertion of the subgoals
produced by a goal 
simplification or backchaining step and we use this ordering to drive
their selection. This eventually determines the point at which
flexible goals are processed. This choice may, on occasion, lead to a
loss of completeness but we believe this to be pragmatically
justifiable. 

\section{The Representation of Lambda Terms}\label{sec:terms}

The abstract interpreter described in the previous section assumes a ready
availability of head normal forms and an immediate access to their
components. In reality, these forms must be
computed. The efficiency of this computation and of the access to the
structures of terms is mediated eventually by the representation
chosen for terms. We discuss the various factors influencing
this choice below, thereby motivating the encoding that
has been used in the {\it Teyjus} system.  Our discussion also
highlights the tradeoffs that are relevant to the representation
question. Lambda terms evolve during computation in a manner that is
difficult to predict statically, making experimentation with actual
implementations a necessary component to quantifying the tradeoffs. An
instrumented version of the {\it Teyjus} system is currently being
used to obtain such an assessment. We indicate some of the
observations from these studies here, leaving a
detailed exposition to other papers \cite{LN02,NX03,LNX03}.

\subsection{The Representation of Bound Variables}\label{ssec:debruijn}
Presentations of lambda terms usually employ a name-based rendition of 
bound variables. When such a representation is used also in an
implementation, it becomes necessary to 
consider the $\alpha$-conversion rule in comparison operations. For
example, a common calculation within higher-order unification is
determining whether the heads of two rigid terms are identical. Thus,
suppose that we desire to unify the terms $\lambdax {y_1}\ldots
\lambdax {y_n} (y_i\app t_1\app \ldots\app t_m)$ and $\lambdax
{z_1}\ldots \lambdax {z_n} (z_i\app s_1\app \ldots\app s_m)$. Term
simplification reduces this task to that of unifying the set
\begin{tabbing}
\qquad\=\kill
\>$\{\langle \lambdax {y_1} \ldots \lambdax {y_n} t_1,
\lambdax {z_1} \ldots \lambdax {z_n} s_1 \rangle, \ldots,\langle
\lambdax {y_1} \ldots \lambdax {y_n} t_m, \lambdax {z_1} \ldots
\lambdax {z_n} s_m \rangle \}$.
\end{tabbing}
\noindent However, a prelude to effecting this transformation is
recognizing that the heads of the two terms match and this clearly
involves a renaming operation under the chosen representation.

If the kind of comparison described above arises often in computation,
it is desirable to use a representation for terms that eliminates the
need for bound variable renaming. A scheme that is suitable from this
perspective is that of de Bruijn \cite{debruijn72}.  Under this
scheme, the connection between binding and bound occurrences of
variables in lambda terms is manifest not 
through names but by using indices at the bound occurrences that count
the number of abstractions in a parse structure of the term up to the
one binding the occurrence. Thus, the term $\lambdax {x}
((\lambdax {y} \lambdax {z} y \app x)\app (\lambdax {w} x))$ is denoted
using the de Bruijn approach by the expression $\lambda\, ((\lambda\,
\lambda\, \#2\app \# 3) \app (\lambda\, \# 2))$, where $\# i$ is the
representation of index $i$. Subterms of a term may have bound
variable occurrences that are free in the local context. An occurrence
of this kind is indicated by an index whose value exceeds the number
of abstractions it is embedded under, as happens in the subterm $(\lambda\,
\# 2)$ of the term considered above. The original de Bruijn
encoding describes also a translation of globally free variable
occurrences to indices. This part of the scheme is, however, not
useful in our context. Globally free variables correspond in our
computational model to variables that can be instantiated. A
characteristic of all the substitutions that we consider for such
variables is that the only unbound variables they contain are ones
that are once again globally free; this property holds, for instance,
of the imitation and projection substitutions discussed in the
previous section. Given this, these variables are best treated, in the
usual logic programming style, as pointers to cells in memory that are
tagged as unbound variables with instantiations being realized
immediately by changing the contents of these cells.

The above discussion indicates a difference in representation and
treatment at a pragmatic level between two kinds of variables that are
similar in the underlying logic. Terminology that distinguishes
between these variables will also be convenient in exposition. We
henceforth use the expression `logic variable' for a  
variable that is globally free, \ie, is not bound by
any explicit abstraction, reserving the terms `bound variable' and
`free variable' for those variables that may be bound or free in a
local context but that are ultimately captured by an abstraction and
hence represented by a de Bruijn index.

The de Bruijn representation solves the problem mentioned at the
outset. The two terms considered there translate under this scheme
into $\lambda\, \ldots \lambda\, (\# i\app \hat{t}_1\app \ldots\app
\hat{t}_m)$ and $\lambda\, \ldots \lambda\, (\# i\app \hat{s}_1\app
\ldots\app \hat{s}_m$), where, for $1 \leq i \leq m$,
$\hat{t}_i$ and $\hat{s}_i$ are the de Bruijn representations of $t_i$
and $s_i$ respectively.  The heads of the two terms are identical
under this representation and, in general, the check for compatibility
of the heads of two rigid terms in the term simplification phase of
unification becomes a simple identity test.

The de Bruijn notation has another significant benefit in that it
allows the abstractions that appear at the front of terms to be
dispensed with in several situations. Such abstractions are often used
in the unification process to encode the contexts in which to view the
two terms that are to be unified.  When these contexts are identical,
as would be the case under the de Bruijn scheme, they can be left
implicit. To understand the pragmatic impact of this observation,
consider again the task of unifying the terms $\lambda\, \ldots
\lambda\, (\# i\app \hat{t}_1\app \ldots\app \hat{t}_m)$ and $\lambda\,
\ldots \lambda\, (\# i\app \hat{s}_1\app \ldots\app \hat{s}_m)$. This
task can be reduced simply to that of unifying the set 
$\{\langle \hat{t}_1, \hat{s}_1 \rangle, \ldots,\langle \hat{t}_m,
\hat{s}_m \rangle \}$,
\ie, the outer abstractions do not need to be appended to
the front of the argument terms. Term simplification thus takes a form
that is closely related to the first-order version: if the heads of
the two rigid terms being considered are identical, the problem simply
becomes one of recursively unifying their arguments. In contrast to
the situation where the outer abstractions need to be replicated and
added in front of the arguments, this transformation is one that can
be easily implemented in a low-level abstract machine.

Although the de Bruijn notation obviates $\alpha$-conversion in the
determination of equality, renaming or, more precisely, renumbering is
still necessary in the correct realization of $\beta$-contraction. To
understand what exactly is needed, let us consider the reduction of
the term $\lambdax {x} ((\lambdax {y} \lambdax {z} y \app x)\app
(\lambdax {w} x))$ whose de Bruijn representation, as we have seen, is
$\lambda\, ((\lambda\, \lambda\, \#2\app \# 3) \app (\lambda\, \#
2))$. This term reduces to $\lambdax {x}\lambdax {z} ((\lambdax {w} x)
\app x)$, a term whose de Bruijn representation is $\lambda\,\lambda\,
((\lambda\, \#3) \app \#2)$. Comparing the two de Bruijn terms, we
observe the following. First, there may be free variables in the
argument part of a redex and the indices corresponding to
these may have to be renumbered as it is substituted into the body
upon performing a $\beta$-contraction; in the example considered, the
subterm $(\lambda\, \#2)$ is transformed into $(\lambda\, \#3)$ by
this process. Second, $\beta$-contraction
eliminates an abstraction and the indices for variable occurrences
that were free in the body have to be decremented by $1$ to
account for this action; once again, in our example, this is reflected
in the renumbering of the index $3$ in the body of the redex to
$2$. This part of the renumbering work can, however, be realized in
the same structure traversal that carries out the substitution of the
argument into the body of the redex.

Renaming is, of course, also necessary in $\beta$-contraction under a
name-based representation. However, in contrast to the situation under
the nameless scheme, renaming now affects only the body of a
redex. Thus, in $\beta$-contracting the term  
$(\lambdax{x} t_1)\app t_2$, it is necessary to consider renaming only
the variables 
explicitly bound within the subterm $t_1$. Even this kind of renaming
can be avoided if it can be determined that the names of these bound
variables do not clash with those of the free variables in
$t_2$. However determining this requires a traversal of the argument
part of the redex to calculate the set of variables that are free in
it. A more efficient approach, used, for instance, in \cite{AP81}, is
to always rename but, in a manner similar to that suggested for the de
Bruijn case, to fold such renamings into the same structure traversal
that realizes the $\beta$-contraction substitution.  

From this discussion, it becomes clear that the separating factor
between the name-based and nameless treatments of bound variables from
the perspective of implementing $\beta$-contraction is the effort
expended in renumbering the argument parts of redexes under
the latter regime. We believe this effort to be 
small in practice for two reasons. First, actual renumbering can be
often be finessed.  For example, if there are no externally bound
variables in the argument of the redex or if substitution is
not made into a context embedded under abstractions, then renumbering
is actually vacuous. This, in fact, is often the situation under a
popular style of programming in $\lambda$Prolog
\cite{miller91jlc},
and other features of the lambda term representation that we describe
in this section allow such properties to be recognized and
utilized in reduction.  Second, not all the cases where a nontrivial
renumbering needs to be done constitute an extra cost. In general, when a 
term is substituted in, it is necessary also to examine its
structure and possibly reduce it to an appropriate normal form. The
necessary renumbering can, in this case, be incorporated into the same
walk as the one that carries out this introspection. The main drawback
of this approach is that it leads to a loss of sharing in reduction if
the same term is substituted, and reduced, in more than one place since
the required renumbering may be different in each of these contexts.
However, empirical evidence suggests that the actual loss of such
sharing is negligible \cite{LN02}, indicating thereby that any
renumbering can be profitably folded into a required reduction walk.  

In summary, then, the de Bruijn representation of bound variables has
few real drawbacks in realizing $\beta$-contraction and significant
advantages in checking identity modulo $\alpha$-conversion and
implementing higher-order unification. It has been used for this
reason in the {\it Teyjus} implementation and we orient the rest
of our discussion of term representation around it.

\subsection{Encoding Substitutions in Terms}\label{ssec:laziness}
The manner in which substitutions are effected over lambda terms is
critical to the efficiency of implementation of the $\beta$-contraction
operation. A potentially desirable feature in the realization of such
substitutions is the ability to perform them in a lazy fashion. For
example, consider the task of determining whether the (de Bruijn)
terms  
$((\lambda\, \lambda\,\lambda\,\#3\app \#2\app s)\app (\lambda\, \#1))$
and $((\lambda\, \lambda\,\lambda\,\#3\app \#1\app t)\app (\lambda\,
\#1))$ can be unified. We assume that $s$ and $t$ denote arbitrary  
terms here.  We can conclude that these two terms cannot be unified by
observing that they reduce respectively to $\lambda\,\lambda\,\#2\app
s'$ and $\lambda\,\lambda\,\#1\app t'$, where $s'$ and $t'$ are
terms that result from $s$ and $t$ by appropriate substitutions. In
making this determination, we do not actually need to calculate the
results of the substitutions over the terms $s$ and $t$. To
achieve this conservation of effort, however, it is necessary 
that we be able to represent $s'$ and $t'$ as  combinations of $s$ and
$t$ with relevant substitutions. Similarly, consider the reduction of   
a term of the form $((\lambda\, ((\lambda\, t_1)\app t_2)) \app t_3)$ to
head normal form. Let $t'_2$ be the term obtained from $t_2$ by
substituting $t_3$ for the first free variable and decrementing 
the indices of all the other free variables by one. Then, producing
the head normal form involves substituting $t'_2$ and $t_3$ for the first
and second free variables in $t_1$ and decrementing the
indices of all other free variables by two. Each of these substitutions 
involves a walk over the {\it same} structure, \ie, the
structure of $t_1$. It would obviously be beneficial if all these
traversals could be combined into one. The ability to do this depends,
once again, on the possibility of temporarily suspending a
substitution generated by a $\beta$-contraction so that it can later
be composed with other substitutions.

The delaying of substitutions has, in fact, been used extensively in
the implementation of functional programming languages (\eg, see
\cite{CCM87,FW87,HM76}). In these contexts, the necessary delaying
is realized by the simple device of combining a term with
an environment that represents bindings for free variables that
occur in it. When the de Bruijn representation is used, this simple
device is  
adequate only if the overall term is closed and if subterms
embedded within abstractions need not be explored. These
assumptions are acceptable in the implementation of
functional programming languages but, unfortunately, not
in the context of interest to us: the production of the head normal
forms needed during unification may well require the
$\beta$-contraction of redexes embedded within abstractions as well as the
propagation of substitutions under abstractions. In these cases, a
more complicated 
substitution operation needs to be encoded. Thus, suppose that we need
to $\beta$-contract a term of the form $(\lambda\, t) \app s$ that
appears embedded within some abstractions. Now, $t$ might contain
variables that are bound by outside abstractions. If the result of
$\beta$-contracting this redex is to be encoded by the term $t$ and an
`environment', the environment must record not just the substitution
of $s$ for the first free variable in $t$ but also the decrementing of
the indices corresponding to all the other free variables. Similarly, 
imagine that we wish to propagate an environment 
under the
abstraction in a term of the form $\lambda\, t$. If the result is to
be represented by a term of the form $\lambda\, t'$ where $t'$ is
itself encoded as $t$ and an environment, then this environment must
be obtained from the earlier one by `shifting up' the index for the
variables to be substituted for by one and adding an identity
substitution for the variable with index $1$. Further, the indices of
the free variables in the terms that appear in the environment must
themselves be incremented by $1$.

Explicit substitution notations that have been developed in recent
years for the lambda calculus offer a complete treatment of this kind
of encoding of substitutions
\cite{ACCL91,BBLR96,Field90,KR97,NW98tcs}. We outline here a version
of such a notation that we have developed for use specifically in the
implementation of our higher-order language
\cite{Nad99jflp}.\footnote{This notation has also been used in the
  Standard ML of New Jersey compiler \cite{shao98:imp}.} Our
notation builds on the traditional de Bruijn notation by adding a 
new category of terms called a suspension. A suspension represents a
`skeletal' term together with a suspended substitution. Such a term
has the structure $\env{t,ol,nl,e}$, where $t$ is a term, $ol$ and
$nl$ are natural numbers and $e$ is an environment. This suspension
corresponds, intuitively, to a term $t$ 
that used to occur inside $ol$ abstractions but that now appears
within $nl$ of them. In generating the underlying de Bruijn term,
therefore, the bound variables with indices greater than $ol$ have to
have their index values adjusted by the difference between $ol$ and
$nl$. Substitutions for 
the first $ol$ bound variables are, on the other hand, contained in
the environment $e$. 
Conceptually, the elements of such an environment are 
either substitution terms generated by a contraction or are dummy
substitutions corresponding to abstractions that persist in an outer
context. 
However, some renumbering of indices may have to be done at the place
of actual substitution. To encode this renumbering, each element of
the environment is annotated with  the number of remaining
abstractions under which the abstraction relevant to that element
appears. This relative `embedding level' can be used
together with the overall embedding level $nl$ to completely determine
the needed renumbering. 

The syntax of lambda terms in the new notation is given formally 
by the category $\langle T \rangle$ defined by the following rules:  
\begin{tabbing}
\qquad\=$\langle ET\rangle$\ \=::=\ \=\kill
\>$\langle ET \rangle$\> ::= \> $\dum \langle N \rangle\ \vert\
(\langle T \rangle, \langle N \rangle)$\\
\>$\langle E \rangle$\> ::= \> ${\it nil}\ \vert\ \langle
ET \rangle :: \langle E \rangle$\\
\>$\langle T\rangle$\>::=\>$\langle C \rangle \ \vert\ \langle
V \rangle \ \vert  \ \#\langle I \rangle\ \vert\ (\langle
 T\rangle\ \langle T\rangle)\ \vert \ (\lambda \langle T\rangle)\
 \vert\ \env{\langle T\rangle,\langle N \rangle, \langle N \rangle,
 \langle E \rangle}$
\end{tabbing}
\noindent In these rules, $\langle C \rangle$ and $\langle V
\rangle$ represent constants and logic variables, $\langle I
\rangle$ is the category of positive numbers 
and $\langle N \rangle$ is the category of natural numbers. Further,
$\langle E \rangle$ and $\langle ET \rangle$ are to be read as the
categories of environments and environment terms, respectively. Terms
of the form $\env{t,i,j,e}$ must satisfy certain wellformedness
constraints that have a natural basis in our informal understanding of 
their content: viewing the environment $e$ as a list, its length must
be equal to $i$, each element of it of the form $\dum(l)$ must be such
that $l < j$ and each element of the form $(t,l)$ must be
such that $l \leq j$. 

\begin{figure}
\begin{tabbing}
\qquad\=(r8)\quad\=\qquad\qquad\qquad\=\kill
\>($\beta_s$)\> $(\lambda t_1)\app t_2 \lra \env{ t_1, 1,
0, (t_2,0) :: {\it nil}}$\\[5pt]
\> ($\beta'_s$)\> $(\lambda
\env{t_1,ol+1,nl+1,\dum{nl}::e})\app t_2 \lra \env{ t_1,ol+1,nl, 
(t_2,nl) :: e }$\\[5pt]
\>(r1)\> $\env{c,ol,nl,e} \lra c$, provided $c$ is a constant.\\[5pt]
\>(r2)\> $\env{x,ol,nl,e} \lra x$, provided $x$ is a logic variable.\\ [5pt]
\> (r3)\>$\env{\#i,ol,nl,e} \lra \#j$, provided $i > ol$ and $j = i - ol + nl$.\\[5pt]
\> (r4)\>$\env{\#i,ol,nl, e} \lra \#j$, provided $i \leq ol$ and $e[i] = \dum(l)$ and $j = nl - l$.\\ [5pt]
\> (r5)\> $\env{\#i,ol,nl,e} \lra \env{t,0,nl - l,{\it nil}}$,\\
\>\> provided $i \leq ol$ and $e[i] = (t,l)$ and $j = nl - l$.\\ [5pt]
\> (r6)\> $\env{t_1\app t_2,ol,nl,e} \lra
\env{t_1,ol,nl,e}\app \env{t_2,ol,nl,e}$.\\[5pt]
\> (r7)\>$\env{\lambda t, ol, nl, e} \lra \lambda \env{t,
ol+1, nl+1, \dum{nl} :: e}$.\\[5pt]
\>(r8)\>$\env{\env{t,ol,nl,e},0,nl',{\it nil}} \lra \env{t,ol,nl+nl',e}$.\\[5pt]
\>(r9)\>$\env{t,0,0,{\it nil}} \lra t$.
\end{tabbing}
\caption{Rule schemata for rewriting terms in the suspension notation}
\label{fig:rewriterules}
\end{figure}

In addition to the syntactic expressions, the suspension notation
includes a collection of rewrite rule schemata whose purpose is to
simulate $\beta$-contractions. These schemata are presented in
Figure~\ref{fig:rewriterules}. In these rules we use the notation
$e[i]$ to denote the $i^{th}$ element of the environment. Of the rules
presented, the ones labelled ($\beta_s$) and ($\beta'_s$) generate the
substitutions corresponding to the $\beta$-contraction rule on de
Bruijn terms and the rules (r1)-(r9), referred to as the {\it reading
rules}, serve to actually carry out these substitutions. 

The ($\beta'_s$) schema has a special
place in the calculus: it makes possible the
combination of substitutions arising from different
$\beta$-contractions. To understand its use, let us consider the
head normalization of the term $(\lambda\, 
((\lambda\, t_1)\app t_2)) \app t_3$. As the first step in this
process, we might produce the term  
$\env{(\lambda\, t_1)\app t_2, 1, 0, (t_3,0) :: {\it nil}}$.
The substitution may now be percolated inwards using the reading rules
so as to reveal a $\beta$-redex at the top level. This produces the term
\begin{tabbing}
\qquad\=\kill
\>$(\lambda\, \env{t_1,2,1,\dum(0)::(t_3,0) :: {\it nil}})\app
\env{t_2,1,0,(t_3,0)::{\it nil})}$.
\end{tabbing}
\noindent At this point the $\beta'_s$ rule schema is applicable and
using it 
produces the term 
\begin{tabbing}
\qquad\=\kill
\>$\env{t_1,2,0,(\env{t_2,1,0,(t_3,0)::{\it nil}},0) :: (t_3,0) :: {\it nil}}$. 
\end{tabbing}
\noindent Notice that the substitutions generated by contracting the
two $\beta$-redexes have been combined at this point into one
environment and can be performed in a single walk over the structure
of $t_1$.  

In the translation of a suspension, it will eventually be necessary to
substitute the arguments of $\beta$-redexes for bound variable
indices. This operation is carried out in our calculus by instances of
the rule schema (r5). There is, in general, a necessity to renumber
indices in the term being substituted in and this is manifest in the
schema (r5) in the construction of a suitable suspension. The rule
schemata (r8) and (r9) recognize special circumstances relative to such
renumbering. The schema (r9) allows vacuous renumbering
to be eliminated. By so 
doing, this rule facilitates a continued sharing relative to the
substituted term. The schema (r8), on the other hand, permits a
nontrivial renumbering walk to be combined with a walk affecting
substitutions arising out of earlier $\beta$-contractions. Uses of the
schemata (r8) and (r9) can be folded into the application of the
schema (r5) and this is actually done in the {\it Teyjus}
implementation. An interesting aspect of our overall system is that by
utilizing the ($\beta'_s$), (r8) and (r9) schemata within the control
strategy for generating head normal forms that we describe later, it
is possible to eliminate nearly all occurrences of nested
suspensions in practice. This has obvious
consequences with respect to the sharing of substitution walks.  

While there is a case in principle for laziness in performing
substitutions, it is still necessary to determine how this plays out
in real applications. In situations where lambda terms are employed in
an essential way in programming, empirical studies indicate that using
the suspension notation and the rules ($\beta'_s$) and (r8)
judiciously can reduce substitution walks to between a third and an
eighth of what is needed when substitutions are performed eagerly
\cite{LN02}. There is a noticeable reduction in computation time as a
result, up to 35\%---measured over all computations, including
backchaining over logic programming clauses---in some important cases.

\subsection{A Dependence Annotation on Terms}\label{ssec:dependency}
There is a refinement to the suspension notation that can have
practical benefits. This refinement consists of annotating
terms to determine whether or not they contain variables bound by 
external abstractions. Referring to the two categories of terms as
open and closed with obvious connotations, these annotations can be
determined statically for de Bruijn terms as follows. At the atomic
level, de Bruijn indices are open whereas logic variables
and constants are closed. For complex terms, an application is open 
if either its `function' or `argument' part is open and is closed
otherwise, and an abstraction is open
exactly when there is a bound variable occurrence within its scope
that has a (relative) index greater than $1$. Rewrite rules that
transform terms in the course of computation can be modified in a
straightforward way to maintain and propagate these annotations. For
example, if a $\beta$-redex of the form $(\lambda\, t_1)\app t_2$ is
closed, then the suspension $\env{t_1,1,0,(t_2,0)::{\it nil}}$ that is
generated from it must also be closed. Similarly, given a suspension
of the form $\env{t_1 \app t_2, ol, nl, e}$ that is closed, the two
top-level components of the term $(\env{t_1,ol,nl,e}\app
\env{t_2,ol,nl,e})$ that is obtained from distributing the substitution
over the application must be closed. A complete presentation of these
refined rewrite rules and a characterization of their properties may
be found in \cite{Nad99jflp}. 

The cost of maintaining the annotations discussed can be made
small by using suitable low-level devices. In the emulator that is
part of the {\it Teyjus} system, for example, an otherwise unused
low-end bit is employed to indicate the annotation and the
determination and setting of its value is folded into the overall
manipulation of term tags. The advantage of maintaining annotations is
at least twofold. First, the rewriting effort in determining the head
normal form of a given term can be reduced. For example, consider a
term of the form $\env{t,i,j,e}$ where it is known that $t$ is a term
that is not dependent on outside abstractions. Then this term can be
simplified immediately to (a pointer to) $t$. Second, this kind of
simplification can foster a greater sharing of terms and,
consequently, of rewriting steps. Thus, consider, once again, the term
$\env{t,ol,nl,e}$, but this time assuming that $t$ is of the form
$(t_1\app t_2)$ that may possibly be shared with other 
contexts. Attempting to reduce this term to head normal form in a
situation where annotations are not used would result in the
production of the term $(\env{t_1,ol,nl,e}\app \env{t_2,ol,nl,e})$, in 
the process breaking the sharing over $t$. In contrast, with the use
of annotations, the given suspension term will be simplified
immediately to $t$ and the subsequent reduction of $t$ will be shared
with all the other contexts in which it is used. 

An obvious question is if the virtues of annotations are
relevant to realistic computations in our higher-order language. We
mention two situations in which these could be of benefit. In the
first instance, observe that the substitutions that are computed by
the higher-order unification process are actually closed in the sense
discussed. Now, if occurrences of the variables being substituted for
appear embedded within $\beta$-redexes, then the propagation of
reduction substitutions over instantiations of these variables can be
calculated trivially by utilizing annotations. As another example
consider a $\beta$-redex of the form $(\lambda\, t_1)\app t_2$ where
$t_2$ is a closed term as would be the case if this term appears 
statically at the top-level. The contraction of this term yields
the suspension $\env{t_1,1,0,(t_2,0)::{\it nil}}$. The percolation of the 
substitution of $t_2$ over the structure of $t_1$ might eventually
lead to the replacement of a bound variable index by $t_2$. With
reference to the rules in Figure~\ref{fig:rewriterules}, this
replacement would produce a term of the form $\env{t_2,0,l,{\it nil}}$, \ie,
a term that corresponds to $t_2$ with a suitable renumbering of
indices corresponding to free variable occurrences. By utilizing the
fact that $t_2$ is known to be closed, the renumbering can be effected
trivially. Furthermore, the bound variable that $t_2$ needs to be
substituted for may occur in more than one place within $t_1$. In this
case the use of the annotation on $t_2$ will also be responsible for
the preservation of a meaningful sharing opportunity. 

At an empirical level, we have observed that the use of annotations
yields a substantial speedup relative to an eager approach to
propagating reduction substitutions, the reduction in computation time
being over 70\% in several cases \cite{LN02}. While there is still a
payoff from annotations when laziness in substitution and the
combination of substitution walks as described in
Section~\ref{ssec:laziness} are used, these appear to be of a much
smaller kind. Thus, certain optimizations seem to overlap with others
and a better understanding of these interactions is needed.

\subsection{The Implementation of Reduction}\label{ssec:red}
An issue of obvious importance is the order in which various
operations are to be carried out on terms. From the perspective of
unification, the main requirement is that of transforming terms into a
form in which the head is exposed; in particular, the arguments of
terms may be left in the form of suspensions. The idea of head normal
forms has been generalized to the suspension notation and its
relationship to the conventional understanding of this notion has been
explored in \cite{Nad99jflp} as a prelude to its use in
unification. At an implementation level, a strategy that might be used
is one that produces these head normal forms only on demand and that
does this by repeatedly rewriting the leftmost, outermost redex
relative to the rules in Figure~\ref{fig:rewriterules} till such time
that an atomic head is revealed, possibly embedded under some
abstractions.  This strategy is an obvious generalization of the one
used for rewriting $\beta$-redexes towards producing head normal forms
in the usual setting and also has practical advantages: it provides
the basis for delaying substitution walks as discussed in
Section~\ref{ssec:laziness} and the different possibilities for
combining term traversals during substitution and the adjustment of
indices present themselves within it as well-defined choices between
rewrite rules applicable at the same time. We discuss the realization
of this approach within the {\it Teyjus} system further in
Section~\ref{ssec:norm}.

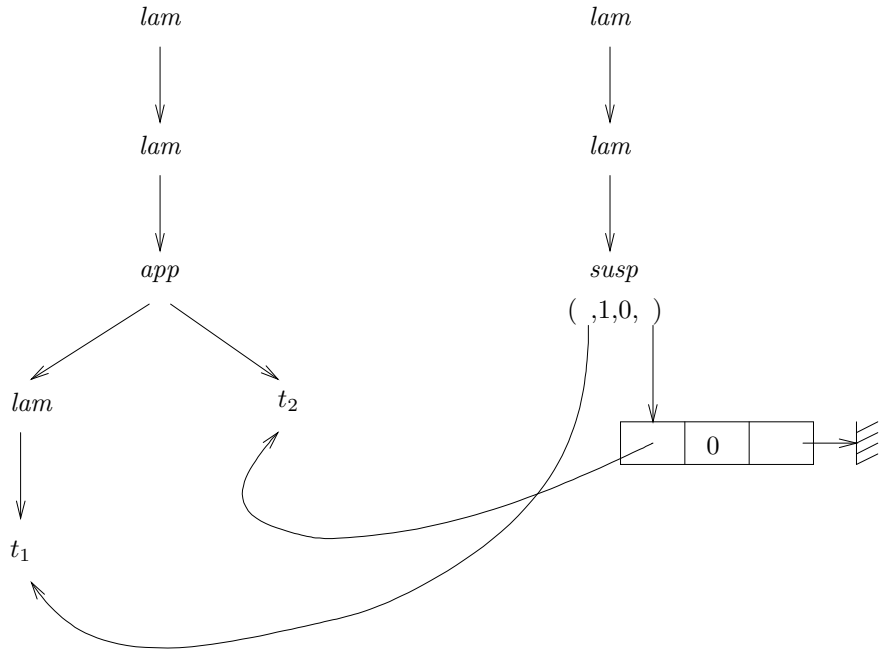
\begin{figure}
\begin{center}
\setlength{\unitlength}{0.00075000in}
\begingroup\makeatletter\ifx\SetFigFont\undefined%
\gdef\SetFigFont#1#2#3#4#5{%
  \reset@font\fontsize{#1}{#2pt}%
  \fontfamily{#3}\fontseries{#4}\fontshape{#5}%
  \selectfont}%
\fi\endgroup%
{\renewcommand{\dashlinestretch}{30}
\begin{picture}(6087,4522)(0,-10)
\path(4725,1297)(4725,1597)
\path(5175,1297)(5175,1597)
\path(5925,1597)(5925,1297)
\path(5925,1522)(6075,1597)
\path(5925,1297)(6075,1372)
\path(5925,1372)(6075,1447)
\path(4275,1597)(5625,1597)(5625,1297)
	(4275,1297)(4275,1597)
\put(4875,1372){\makebox(0,0)[lb]{{\SetFigFont{10}{12.0}{\rmdefault}{\mddefault}{\updefault}0}}}
\path(5925,1447)(6075,1522)
\path(4500,1447)(4499,1447)(4497,1446)
	(4493,1444)(4487,1441)(4478,1437)
	(4467,1431)(4452,1424)(4434,1415)
	(4412,1405)(4387,1393)(4359,1380)
	(4327,1365)(4293,1349)(4256,1332)
	(4216,1313)(4174,1294)(4131,1275)
	(4085,1254)(4038,1234)(3990,1213)
	(3941,1192)(3890,1170)(3839,1149)
	(3786,1128)(3733,1107)(3679,1086)
	(3623,1066)(3566,1045)(3509,1025)
	(3449,1005)(3388,985)(3326,966)
	(3262,947)(3196,928)(3128,910)
	(3060,893)(2990,876)(2920,861)
	(2850,847)(2765,832)(2683,819)
	(2607,809)(2537,801)(2474,794)
	(2418,789)(2369,785)(2326,783)
	(2289,781)(2257,780)(2229,780)
	(2206,780)(2186,781)(2168,782)
	(2152,783)(2138,784)(2123,787)
	(2109,789)(2094,792)(2078,795)
	(2060,799)(2040,804)(2018,809)
	(1993,815)(1965,823)(1935,832)
	(1902,842)(1866,854)(1830,868)
	(1793,884)(1758,902)(1725,922)
	(1693,947)(1669,974)(1650,1003)
	(1637,1031)(1630,1060)(1627,1090)
	(1629,1119)(1634,1148)(1642,1178)
	(1653,1207)(1667,1236)(1682,1265)
	(1700,1294)(1718,1323)(1737,1351)
	(1757,1377)(1776,1403)(1794,1426)
	(1812,1448)(1827,1467)(1841,1483)
	(1852,1496)(1861,1506)(1875,1522)
\path(1818.557,1411.936)(1875.000,1522.000)(1773.402,1451.446)
\path(4050,2272)(4050,2271)(4050,2269)
	(4050,2266)(4050,2261)(4050,2254)
	(4050,2244)(4050,2231)(4050,2215)
	(4049,2197)(4048,2175)(4048,2151)
	(4046,2123)(4045,2093)(4042,2061)
	(4040,2026)(4036,1989)(4032,1951)
	(4027,1911)(4022,1869)(4015,1827)
	(4007,1783)(3998,1739)(3988,1694)
	(3976,1648)(3963,1601)(3947,1555)
	(3930,1507)(3911,1459)(3890,1410)
	(3866,1361)(3840,1311)(3810,1260)
	(3777,1208)(3741,1155)(3701,1102)
	(3657,1048)(3609,993)(3556,937)
	(3500,882)(3439,826)(3375,772)
	(3314,724)(3251,678)(3187,634)
	(3124,592)(3062,553)(3002,517)
	(2945,484)(2889,453)(2837,425)
	(2788,400)(2741,377)(2698,357)
	(2658,338)(2620,322)(2585,307)
	(2552,294)(2521,283)(2492,273)
	(2464,264)(2438,255)(2412,248)
	(2387,241)(2362,234)(2338,228)
	(2313,222)(2287,216)(2260,210)
	(2232,203)(2202,197)(2170,189)
	(2136,182)(2099,173)(2060,165)
	(2017,155)(1971,145)(1922,134)
	(1870,123)(1814,111)(1754,98)
	(1692,86)(1626,73)(1558,61)
	(1488,50)(1417,39)(1346,30)
	(1275,22)(1187,15)(1103,12)
	(1024,13)(951,16)(882,22)
	(819,30)(760,40)(706,53)
	(656,67)(610,83)(567,100)
	(528,118)(491,138)(456,158)
	(424,180)(394,202)(366,224)
	(340,247)(315,270)(293,293)
	(271,315)(252,337)(234,357)
	(219,377)(205,395)(192,411)
	(182,426)(173,438)(166,448)
	(160,456)(156,463)(150,472)
\path(241.526,388.795)(150.000,472.000)(191.603,355.513)
\put(900,2572){\makebox(0,0)[lb]{{\SetFigFont{10}{12.0}{\rmdefault}{\mddefault}{\updefault}{\it app}}}}
\put(4050,3472){\makebox(0,0)[lb]{{\SetFigFont{10}{12.0}{\rmdefault}{\mddefault}{\updefault}{\it lam}}}}
\put(900,3472){\makebox(0,0)[lb]{{\SetFigFont{10}{12.0}{\rmdefault}{\mddefault}{\updefault}{\it lam}}}}
\put(4050,4372){\makebox(0,0)[lb]{{\SetFigFont{10}{12.0}{\rmdefault}{\mddefault}{\updefault}{\it lam}}}}
\put(900,4372){\makebox(0,0)[lb]{{\SetFigFont{10}{12.0}{\rmdefault}{\mddefault}{\updefault}{\it lam}}}}
\put(0,1672){\makebox(0,0)[lb]{{\SetFigFont{10}{12.0}{\familydefault}{\mddefault}{\updefault}{\it lam}}}}
\path(75,1522)(75,922)
\path(45.000,1042.000)(75.000,922.000)(105.000,1042.000)
\path(5550,1447)(5925,1447)
\path(5805.000,1417.000)(5925.000,1447.000)(5805.000,1477.000)
\path(975,2422)(150,1897)
\path(235.133,1986.735)(150.000,1897.000)(267.346,1936.115)
\path(1125,2422)(1875,1897)
\path(1759.488,1941.239)(1875.000,1897.000)(1793.896,1990.392)
\path(4200,3322)(4200,2797)
\path(4170.000,2917.000)(4200.000,2797.000)(4230.000,2917.000)
\path(1050,3322)(1050,2797)
\path(1020.000,2917.000)(1050.000,2797.000)(1080.000,2917.000)
\path(1050,4222)(1050,3697)
\path(1020.000,3817.000)(1050.000,3697.000)(1080.000,3817.000)
\path(4200,4222)(4200,3697)
\path(4170.000,3817.000)(4200.000,3697.000)(4230.000,3817.000)
\path(4500,2272)(4500,1597)
\path(4470.000,1717.000)(4500.000,1597.000)(4530.000,1717.000)
\put(1875,1672){\makebox(0,0)[lb]{{\SetFigFont{10}{12.0}{\familydefault}{\mddefault}{\updefault}$t_2$}}}
\put(4050,2572){\makebox(0,0)[lb]{{\SetFigFont{10}{12.0}{\familydefault}{\mddefault}{\updefault}{\it susp}}}}
\put(0,622){\makebox(0,0)[lb]{{\SetFigFont{10}{12.0}{\familydefault}{\mddefault}{\updefault}$t_1$}}}
\put(3900,2272){\makebox(0,0)[lb]{{\SetFigFont{10}{12.0}{\familydefault}{\mddefault}{\updefault}(\,\ ,1,0,\,\    )}}}
\end{picture}
}

\end{center}
\caption{Non-destructive reduction of lambda terms}
\label{fig:internal}
\end{figure}

Another issue to consider is whether to implement the rewriting of
terms in a destructive or non-destructive manner. To understand 
the tradeoffs involved, let us consider the reduction of the term
$\lambda\, \lambda\, ((\lambda\, t_1)\app t_2)$, in which 
$t_1$ and $t_2$ are arbitrary terms, to head normal form.
Anticipating a discussion of internal representations, we may depict 
terms by graphical structures. Each term in such a representation
translates to a node labelled with its category and containing its
fixed length parts and pointers to relevant subterms and
environments. Assuming such a visualization, the internal structure of
the term of interest may be shown by the graph in the
left half of Figure \ref{fig:internal}. Now, this term has a
redex, given by the subterm $(\lambda\, t_1)\app t_2$, that
has to be $\beta$-contracted in producing a head normal form.  If a
destructive implementation of reduction is used, then this rewriting
step will be effected by replacing the redex in-place by the 
term $\env{t_1,1,0,(t_2, 0)::{\it nil}}$.\footnote{The representation of
different kinds of terms generally require different amounts of space
in an concrete realization. In this case, a destructive change may
be achieved by using a special kind of term that serves as a
`reference' to a value stored elsewhere.}
The consequences of this replacement will be felt immediately in all
places where the redex appears as a subterm. If the
$\beta$-contraction is done non-destructively, on the other hand, the
subterm would be left intact but a new suspension term would be
returned. To obtain the effect of this rewriting step in the overall
context, it would now be necessary to replicate around the suspension
the structure of the term within which the redex is embedded.
Thus, a non-destructive implementation of the
$\beta$-contraction operation would eventually have to produce the
structure shown in the right half of Figure \ref{fig:internal}. 

Based on the above understanding, a destructive implementation
appears to be an obvious choice in a context where 
reduction is deterministic, \ie, where future
events will not require rewriting steps to be undone. The in-place
replacement obviates a copying of the embedding context. Furthermore,
it is only such a replacement that admits of any possibility of
sharing in 
reduction; the effect of replacing $(\lambda\, t_1)\app t_2$ with
$\env{t_1,1,0,(t_2,0)::{\it nil}}$ will be felt in other contexts only if
the term is changed physically at the place where these point to. A
non-destructive implementation actually has an additional cost that is
inhibiting. Consider, for instance, an attempt to head normalize a
term that is already in head normal form but that is not known
statically to be in this form. Since copying of structure is needed in
some cases, a naive implementation might simply replicate the
structure of the term even when its subparts are unchanged. However,
this is undesirable: a mere `look-up' should not cause a new structure
to be created. This kind of a copying can be avoided by putting
explicit checks into the normalization procedure to determine when
copying is necessary.  This incurs a time penalty that does not arise
in a destructive implementation.

Our interest is, of course, eventually in a context where backtracking
over reduction computations may be necessary. As a specific example, a
$\beta$-redex may manifest itself in a term as a result of a
substitution for a variable that may have to be repealed at a later
point. In this situation, a destructive implementation has the drawback
that the in-place changes may have to be trailed to facilitate a
subsequent resetting of state. A interesting point to note is that
the form of trailing that is needed here is one that saves old values
of cells and not simply the pointers to affected cells as in
conventional Prolog implementations. The efficiency of such an
implementation can obviously be improved by mechanisms that detect
redundancy in trailing. The simplest method that can be used for this
purpose is one that compares the location in heap of the term being
changed with the most recent heap backtrack point to decide the
necessity for trailing. Controlled forms of eagerness that push
necessary rewriting steps to before the setting up of choice points
are also useful. Another aspect that bears careful investigation is
the possibility of committing to heap a cascade of reduction
steps, such as those corresponding to a $\beta$-contraction and the 
propagation of the substitution it generates, only at the very end,
thereby obviating the retraction of intermediate steps. The present
version of the {\it Teyjus} system employs a graph based, 
destructive implementation of reduction and, as such, provides an
effective vehicle for experimenting with these various
possibilities.\footnote{Such experimentation with reduction
  strategies has actually been carried out since the preparation of
  this paper. Details may be found in \cite{NX03} and \cite{LNX03}.}

The implementation of logic programming languages have, in the past,
consider two different methods for the treatment of terms appearing in
program clauses. In the structure sharing approach, these
are represented as a combination of a fixed structure and bindings for
variables \cite{War77}. In the more popular structure copying approach,
entirely new copies of these terms are created through compiled code
in each instance of use \cite{War83}. The destructive
implementation of reduction is compatible only with this structure
copying approach and we therefore assume its use in later
sections. 

\subsection{Internal Representation}\label{ssec:intrep}
The final issue that we consider is the low-level representation of
terms that embodies the various mechanisms that we have described.
The most natural such encoding is the one that uses a cell
bearing a tag that indicates the relevant syntactic category for each
term and that is complemented by additional cells containing
information about further components. In the case that the dependency
annotations discussed in Section~\ref{ssec:dependency} are used, it is
best if the information they provide can also be accessed
independently of the term category. In the emulator underlying the {\it
Teyjus} implementation, this purpose is achieved by reserving the
low-end bit of the tag bearing cell for these annotations. 

The information that needs to be provided in addition to the term
category depends, of course, on the category of the term. 
If this is a constant or a bound variable, all that is needed is a
pointer to the descriptor for the constant or the index, and this can
be folded into the cell bearing the tag. In the case of a
logic variable that is as yet uninstantiated, the contents of the cell
are unimportant and the instantiation of such a variable is realized by
changing the contents of this cell to correspond to one of the other
term categories. An abstraction cell 
must contain a pointer to the body of the abstraction. 
A suspension term requires the maintenance of its two
indices, a pointer to the skeletal term and a pointer to its
environment. An environment can be represented as a list and,
in this form, admits of considerable sharing. 

The representation of the final category of terms, namely,
applications, requires more care. The most natural, and
perhaps the conceptually clearest, approach is to utilize the curried
structure, rendering each application into a pair of pointers to its function
and argument parts. Unfortunately, this kind of rendition incurs a high
cost in the most common form of access to terms. The objective with
terms is typically to get to the heads of their head normal
forms. Further, operations such as term simplification in unification
are best realized if the arguments in a head normal form are available
as a vector. Suppose that we have a term that at compile time has the
structure $\lambda\,\ldots\,\lambda\, (h \app t_1 \ldots t_n)$. If a
curried representation is used for this term, $n$
applications will have to be traversed before the head is
reached and a vector of arguments will also have to be
explicitly constructed in the course of this descent under application
structure.  

An alternative encoding of application that is reminiscent of the
treatment of terms in conventional logic programming
implementations is to translate it into a structure containing three
components: a function part, a pointer to a {\it vector} of arguments
and an `arity' that indicates the size of the arguments vector. Such a
representation has especially nice properties when the program at hand
is a first-order one. In this case the top-level structure of every
compound (application) term that is encountered during computation is
already available at compile time. Thus, the head normal forms of
these terms are available without any reduction calculations and
the described representation allows a quick determination of this fact
as well as an immediate access to the functor and arguments
parts. These appear to be important properties since efficiency over
first-order like computations is significant even to a higher-order
logic programming language \cite{MP92}. 

Our low-level representation for terms comes close to the 
one generally employed for first-order terms with the described
encoding of applications. However, there are still differences that
should be mentioned. One difference concerns the specific structure
chosen for compound terms. In the first-order case, internal nodes in
the tree representation of terms cannot change. This fact can be
exploited to fold the functor and arguments parts into one vector and,
thereby, to reduce compound terms to a single pointer. A similar
optimization is not possible in the higher-order context. The
contraction of a $\beta$-redex, for instance, transforms an $n$-fold
application into an $(n-1)$-fold one and there must be sufficient
flexibility in the encoding of terms to capture this situation. 
The other difference lies in the registration of destructive
changes for the purposes of backtracking. First-order terms 
evolve during computation only by virtue of bindings for logic
variables. By picking a uniform representation for such variables, 
the state prior to such a change can be recorded simply by retaining a
pointer to the cell for the corresponding variable. This kind of
optimization is not available in the higher-order context when
reduction is implemented destructively and, as we have already noted,
the original value of the modified cell needs also to be remembered in
order to resurrect the previous state.

We have up to this point not considered explicitly
the fact that the terms of interest to us have types
associated with them. These types have a twofold role in the language
\cite{NP91}. At one level, they serve to limit the set of acceptable
programs. At another level, they participate in the computational
mechanism of the language; this role is apparent from the manner in
which types determine the imitation and projection substitutions that
are to be generated for a flexible-rigid pair. The first function of
types is relevant to compilation but does not affect the
execution of a program and so does not have a bearing on runtime
representations. As for the second 
purpose, we observe that it is sufficient to maintain types
with only the constants and logic variables appearing in lambda
terms. Maintaining such annotations is also necessary: 
the types of logic variables are needed for both the imitation
and projection substitutions and the types of constants are needed in
determining the imitation substitutions. 

The need to maintain types adds a extra component---a pointer to
a type---to the representation of logic variables. The
representation of constants is unchanged since the type information
can be combined with the other data comprising their descriptors. 
While we do not discuss this issue in detail here, in the
presence of polymorphism, types are best represented by pointers to a
type `skeleton' and a type environment \cite{KNW92}. 
The treatment of polymorphism thus adds an extra cell to the
representations of constants and logic variables.

\section{Runtime Support for Higher-Order Unification}\label{sec:runtime}

We now consider the task of supporting the enhanced notion of
unification present in our language. The
problems that have to be dealt with are threefold. First, it is
necessary to consider the normalization of terms during
execution. Second, states in our abstract interpreter are given also
by disagreement sets and an efficient method for maintaining such sets
explicitly is needed. Finally, higher-order unification has a
branching character, a facet we realize through depth-first search
with backtracking. In implementing this approach, it is necessary to
identify the important components of state that need to be remembered
and also to describe suitable encodings for such information. We
discuss these various issues below and we describe approaches to
accounting for them within an actual implementation.

\subsection{Normalization of Terms}\label{ssec:norm}

The simplification operation and the postulation of substitutions
within unification depend on terms being presented in (a generalized)
head normal form. Terms can arise during computation that are not in
this form; consider, for instance, the structure of a term after a
substitution dictated by imitation or projection has been made for a
variable of function type that appears in it. Mechanisms for
normalizing terms are therefore needed as also is a protocol for
deploying these at points where head normal forms are desired. 
As discussed in Section~\ref{ssec:red}, a strategy that rewrites the
leftmost, outermost redex at each stage is a natural one to use for
head normalization. The suspension notation allows the substitution
generated by a $\beta$-contraction to be treated as a truly atomic
operation and thereby facilitates an iterative, stack based
realization of this strategy. Such an approach is embedded in the
implementation of normalization within the {\it Teyjus} system. We
sketch this component of the system below as a prelude to
explaining its use in the overall computation scheme. A more detailed
description of the reduction procedure may be found in \cite{Nad98esw}.  

The {\it Teyjus} implementation of head normalization actually uses
two stacks called the structures list or {\it SL} stack and the
applications stack, the latter facilitating a destructive 
realization of reduction. Both stacks store references to terms and
can share a common space in an abstract machine, with their tops
growing towards each other. The reduction procedure looks at the term
pointed to by the top of the {\it SL} stack and the value in a global
register {\it NUMARGS} to determine its next step. At the outset, a
reference to the term to be reduced is placed on the top of the {\it
SL} stack and the {\it NUMARGS} register is set to $0$.  The main
actions of the procedure are dependent on the term referenced by the
top of the {\it SL} stack having a non-suspension structure. For this
reason, if this term is a suspension, the first task becomes that of
exposing such a form for it.  This objective is achieved immediately
using one of the reading rules in Figure~\ref{fig:rewriterules} when
the skeleton is itself not a suspension. Otherwise a non-suspension
form must be exposed first for the skeleton and a simple iterative
process that begins by placing a reference to this skeleton on the top
of the {\it SL} stack serves to realize this.  Eventually, when a
non-suspension form is exposed, if the top of the {\it SL} stack is a
reference to an application, this reference is recorded in the
applications stack and is replaced in the {\it SL} stack by a sequence
of references to its ``operand'' and ``operator'' parts, with the {\it
NUMARGS} register being incremented by the number of operands. If the
top of the {\it SL} stack contains a reference to an abstraction, the
action taken depends on the value in the {\it NUMARGS} register. If
this is $0$, the {\it SL} stack reference is replaced by one to the
body of the abstraction. Otherwise, a leftmost, outermost
$\beta$-redex has been found and needs to be contracted. This action
is realized by popping the top two items on the {\it SL} stack, using
them to construct a suitable suspension a reference to which is pushed
onto the {\it SL} stack, destructively updating the application
available from the top of the applications stack and, finally,
decrementing the {\it NUMARGS} register by $1$ to account for the
disappearance of an argument. The final possibility for the top of the
{\it SL} stack is that it is a reference to an atomic term, \ie, one
that is a constant or a bound or free variable. This situation signals
that a head normal form has been found and hence terminates the
overall process. 

To understand the integration of the head normalization procedure 
into the larger computational framework, suppose that it is invoked
with a term that can be reduced to the form $\lambdax {x_1}
\ldots \lambdax {x_n} (h \app t_1 \app \ldots \app t_m)$, where $h$ is
a constant or variable. When the procedure is finished, the {\it SL}
stack will contain, in consecutive locations from the top, references
to the de Bruijn representation of $h$ and the 
suspension representations of terms $s_1, \ldots, s_m$ that are
$\beta$-convertible to $t_1, \ldots, t_m$. This kind of access to the
body of the term is particularly convenient for the other operations
required within unification. First of all, the heads of terms and
their status, whether rigid or flexible, is easily
determined. Further, assume that the simplification operation is to be
applied to two terms $t$ and $r$ whose head normal forms have 
identical binders. The head normalization procedure can, in this case,
be invoked to lay out the bodies of these two terms in different
segments of the {\it SL} stack. Then, if the two terms are rigid and have
identical heads, the terms out of which new disagreement pairs have to
be formed appear at the same displacement from different starting
locations, thereby facilitating an iterative structure to further
processing. The availability of the arguments of the head normalized
term in contiguous locations turns out also to be important to the
compilation model that we discuss in Section \ref{sec:abstmachine}.

Our description  of the term simplification process above assumes that
the lengths of binders of the two terms to be unified are identical. This
situation may actually not hold automatically  
but, rather, may have to be achieved at the required points in
computation by using the $\eta$-conversion rule. 
A few simple changes to our normalization routine suffices for making
the requisite adjustments. Thus, suppose that we are comparing
two terms that have as head normal forms $\lambdax {x_1} \ldots
\lambdax {x_n} (c\app t_1\app \ldots \app t_l)$ and $\lambdax {x_1}
\ldots \lambdax {x_m} (c'\app s_1\app \ldots \app s_k)$. Our
procedure can easily record the values of $m$ and $n$ when producing
these forms. Now suppose that $n$ is greater than $m$ and that $c'$ is
identical to one of $x_1,\ldots,x_m$. Then the effect of adjusting the
binder length of the second term to $n$ on its head in a suspension
representation of the terms can be captured simply by adding $n-m$ to
the index value corresponding to $c'$. The changes to the first $k$
arguments of this term under such an adjustment are also
straightforward to capture: if $s'_i$ is the term in suspension 
notation corresponding to $s_i$, the desired adjustment is
encapsulated in the term $\env{s'_i,0,j,{\it nil}}$ where $j =
n-k$. Finally, new arguments need to be added to the term, but this is
particularly easy to do on-the-fly, they being just the de Bruijn
indices $\#(n-m), \ldots, \#1$. 

The normalization process requires the creation of new structures for
terms at various points in its execution. These terms are best
allocated on the heap in a WAM-like model and this is, in fact, what
is done within the {\it Teyjus} implementation. 

\subsection{Explicit Representation of Disagreement Sets}\label{ssec:disset}

Disagreement sets arise in principle even in the context of
first-order unification. However, typical implementations of this
operation avoid the explicit treatment of such sets by utilizing a 
recursive, depth-first processing of the subparts of the two terms
that are to be unified. Careful attention to the order in which
subterms are processed is known to make a substantial difference to
the worst case behaviour \cite{MM82}, but the `pathological' cases
seldom seem to arise in practice. Given this, flexibility in choosing
the next pair of (sub)terms is usually sacrificed for a simpler
processing structure.

Two properties of first-order unification are actually critical to
adopting the approach that treats disagreement sets implicitly: its
decidability and the existence of most general unifiers. Neither of
these properties carry over to the higher-order context. While it is
still possible to use a recursive process that explores unifiers for
subterms in a depth-first fashion, basing the unification computation
entirely on such an approach appears pragmatically undesirable. In
particular, in a situation where choices have to be made in
substitutions, it appears best to bring all available constraints to
bear on making them. Thus, suppose that it is necessary to
unify two terms of the form $(f \app t_1 \app \ldots \app t_n)$ and
$(f\app s_1 \app \ldots \app s_n)$, where $f$ is a constant (function)
symbol and, for $1 \leq i \leq n$, $t_i$ and $s_i$ are arbitrary
terms. We may well attempt to do this by unifying the pairs $\langle
t_1,s_1\rangle,\ldots,\langle t_n,s_n \rangle$ in sequence. Now, in
the course of unifying the pair $\langle t_1, s_1
\rangle$, it may be necessary to pick one of several substitutions for
a variable $x$. This variable may appear in other pairs of subterms as
well and some of the substitution choices for $x$ may render these
pairs non-unifiable. Using this information, at the very least,
curtails the search. In particular cases, this may even 
make a difference between finding and not finding a unifier: some
choices of substitution for $x$ that are ruled out by their effects on
other pairs may lead to a unending search when the pair $\langle
t_1, s_1 \rangle$ is considered in isolation. 

A better approach to finding unifiers for a disagreement set, then,
appears to be the following. At any point in the computation, we
select a pair from the set and proceed to search for a unifier for
only this pair till such time that more than one possibility needs to
be explored. At such a point, we pick a
possible substitution and examine its effect on the rest of the
disagreement set before proceeding further. Implementing this approach
clearly requires an explicit representation and manipulation of
disagreement sets within the unification process. Actually, 
such disagreement sets may even have to be carried {\it across} invocations
to unification: repeated applications of the unification step
described in Section \ref{sec:interp} may reduce a disagreement set to
a form in which, while being nonempty, it contains only
flexible-flexible pairs and, as we have noted already, it is best to
suspend the processing of such a set till further constraints are
imposed on it through backchaining steps. 

Given that we need to maintain disagreement sets explicitly, an
important question becomes that of the structure their representation
should take. The following considerations are relevant to this issue:

\begin{enumerate}

\item These sets evolve incrementally during computation. In
particular, changes result from adding new pairs to an existing set or
by effecting substitutions that modify only some pairs leaving the
others unchanged. Thus, a scheme that allows the representation of a
new disagreement set to reuse that of unchanged portions
of the set from  which it arises might be the preferred one in
practice. 

\item For efficiency in backtracking, it should be possible to rapidly 
recreate disagreement sets that were in existence earlier. This
becomes a pertinent issue if the kind of sharing described in (1) is
realized through destructive changes. 

\end{enumerate}

A representation for disagreement sets that suffices for meeting the
above requirements is one based on doubly linked lists the elements
of which are pairs of pointers to the terms constituting the
pairs in the set. Given that these sets arise in the
course of backchaining or clause invocation and finally disappear in
the event of backtracking, the lists representing them are naturally
allocated in the heap in a WAM-like setting.  The need to examine
disagreement sets during computation requires that the beginning of
the lists representing them be recorded in machine states. A special
register that we refer to as the live list or {\it LL} register and
that contains a reference to the first element in the list at
each execution point can serve this purpose. Now, there are
two ways in which a disagreement set might change during
computation. First, term simplification may require some element of
the set to be removed and new pairs corresponding to subterms that
need to be unified to be added to the set. The removal of a pair is
realized in this setting by changing the `after' and `before' pointers
of the elements on either side of it in the list representation. A
subsequent re-inclusion of the removed pair into the set
can be effected easily if a reference to it is maintained,
something that can be done through a (properly annotated) entry in the 
trail stack. The addition of new pairs is also simple: entries for
these can be created on the heap and added to the beginning of the
live list. The second way in which a disagreement set may change is
through a backtracking operation. To support this action, it is
necessary also to store the contents of the {\it LL} register in
choice points at the time of their creation. The relevant disagreement
set can then be resurrected by utilizing information in the trail
stack to restore deleted pairs and using the old value of the {\it LL}
register to remove the pairs added beyond the point being backtracked
to. 

It is in principle possible to perform all the processing within the
term simplification phase of unification using only the heap and the
live list. However, judiciousness should be exercised in utilizing
the heap since space allocated in it is reclaimed only through
backtracking. With this in mind, we  
observe that when a rigid-rigid pair is encountered during term
simplification, the processing can be applied recursively to the
subterms and additions to the heap and the live list need not take
place till a flexible-flexible or a flexible-rigid pair is
encountered. As a specific example consider the following query
involving the {\it mapfun} predicate from Example~\ref{ex:mapfun}: 
\begin{tabbing}
\qquad\=\kill
\>${\it mapfun}\app (a :: b :: {\it nil}) \app G \app (g\app a\app a) ::
(g\app a\app b) :: {\it nil})$
\end{tabbing}
\noindent Using the idea just described, when applying a backchaining
step to this query relative to the second clause for {\it mapfun}, the
addition to the heap of only the pairs in the set
\begin{tabbing}
\qquad\=\kill
\>$\{\langle X, a \rangle, \langle L1, b::{\it nil} \rangle, \langle F, G \rangle,
\langle (F\app X), (g\app a \app a) \rangle, \langle L2, (g\app a \app b)::{\it nil}
\rangle\}$
\end{tabbing}
\noindent need be considered. When one of the terms in a disagreement pair
is known statically, this kind of processing can be realized through
special instructions and a compilation process similar to that used in
the first-order case, and we discuss this matter in greater detail in
the next section. However, the two rigid terms in a pair can sometimes
arise dynamically and term simplification has in this case to be
carried out in `interpretive' mode. In the context of a virtual
machine based implementation, a special pushdown list in combination
with an iterative code fragment can be used to realize this
computation. 

There are certain forms of disagreement pairs for which most general
unifiers can be immediately identified. An example of this kind
arises from first-order unification: given a pair of the form 
$\langle X, t \rangle$ where $X$ is a variable of atomic type and $t$
is a term in which $X$ does not appear, all unifiers for the pair must
be instances of the unifier that substitutes $t$ for $X$.
Alternatively, if $X$ does occur in $t$, failure in unification can be
registered immediately. This observation can be generalized to the 
higher-order context. Given a pair of the form $\langle
\lambdax {x_1} \ldots \lambdax {x_n} X, \lambdax {x_1} \ldots \lambdax
{x_n} t \rangle$, where $X$ has an arbitrary type, $X$ can be bound to
$t$ and this pair can be removed from the set provided neither $X$ nor
the variables in $\{x_1,\ldots x_n\}$ appear in $t$; interestingly, the
verification of the proviso is simplified by the use of the de Bruijn
notation. The occurrence of $X$ (or of the other variables) in $t$ on
the other hand does not by itself signal failure in the higher-order
case. However, the `occurs-check' from the first-order case can be
generalized to a `rigid path check' 
that detects the impossibility of unification in some cases and that
simplifies the search for unifiers in other cases by binding $X$ to a
term that represents an initial `section' of $t$ and by adding pairs to
the disagreement set to represent the remaining constraints on
unifiers. Some flexible-flexible pairs, such as $\langle F, G \rangle$
in the example above, can also be solved by this process. Using
observations such as these reduces the need to consider 
the general imitation and projection substitutions and hence also the
attendant bookkeeping steps. In the case of the ${\it mapfun}$ query, the
disagreement set can, in fact, be quickly reduced to $\{\langle (F\app a),
(g\app a\app a) \rangle \}$ by these means. Significantly, first-order
unification can be solved immediately using these observations.
Empirical studies indicate that a large number of 
the unification problems that arise even in the higher-order context
fall into this category \cite{MP92}, suggesting the general importance 
of incorporating these observations into an
implementation.\footnote{The applicability of this first-order like
processing in the higher-order case is dependent on variables being
preserved in an `$\eta$-reduced' form through the compilation
process. The procedure described in \cite{DHKP98} for unification
of what are known as higher-order patterns \cite{miller91jlc}
gives up this property. This is a surprising choice, especially since
it is not dictated by the theory: under it, even the
pair $\langle F, G\rangle$ gets converted into a form that
requires an `inverting' substitution to be computed. The only possible
benefit for this is that variables need never be dynamically
`$\eta$-expanded.' However, this is {\it rarely}, perhaps {\it never},
required in practice. The direct use of explicit 
substitutions also does not seem to have the practical benefits in this
subcase that it has for full higher-order unification \cite{DHK00},
and it possibly has some drawbacks. We are attempting to quantify
these remarks in ongoing research.}

The appropriate time to consider such substitutions is during 
the term simplification phase. Doing this and also being
conservative in the additions to the heap now calls for the use of two
pushdown lists. The general scheme works as follows. The
simplification of a disagreement set 
proceeds as before with the use of the first pushdown list, except
that the process may now also involve making bindings to variables.
When the process `bottoms out' with a flexible-flexible or
flexible-rigid pair, this is pushed onto the top of the second
pushdown list instead of the heap. When all the pairs in the
original disagreement set have been simplified, it is checked whether
any bindings were made in the course of simplification. If no bindings
were made, the pairs in the second pushdown list are transferred to
the heap and included in the live list. If, on the other hand, any
bindings were made, the simplification process is repeated with the
disagreement set being given now by the live list and the pairs in the
second pushdown list and the roles of the two pushdown lists
being reversed. 

\subsection {Recording Branch Points in Unification}

A depth-first approach to exploring alternatives in unification 
requires that information be recorded at branch points that
is sufficient for recreating state and for determining the remaining
possibilities upon backtracking. The state information can actually be
factored into two conceptual kinds: that which pertains to clause
usage in the backchaining model of computation and that which 
relates to the unification problem, such as the disagreement set and
the pair of flexible and rigid terms that are under consideration at a
particular juncture. Now, the approach that we have proposed is one
that solves unification problems to the extent possible before
contemplating further backchaining steps. Under this strategy,  
in situations where genuine higher-order unification is involved, a
sequence of branch points are likely to be generated for which the
state insofar as it pertains to clause usage is identical. Thus, if
this information is represented separately from the rest of the
backtracking data, it becomes possible to share it across more than
one branch point. 

In light of the above observations, we propose to record
information relevant to branching in unification in two layers that we 
refer to, respectively, as the shared part and the variable
part. Assuming a WAM style 
compilation model, the most appropriate juncture at which to consider
genuine higher-order unification is right after a compiled form of
term simplification akin 
to first-order unification has been carried out relative to the head
of the clause and before an attempt is made to solve the clause
body. At this stage, if the {\it LL} register indicates a nonempty
disagreement set, the first action would be to create the shared part
of the unification branch point records that stores clause usage
information. More specifically, this part would record at least the
following state data: 
\begin{itemize}
\item The program pointer that determines the instruction to be
executed upon successful completion of this phase of unification.

\item A pointer to the most recent environment record.

\item The continuation pointer; this is relevant in the case of
clauses with an atomic body for which an environment record does not
exist containing this information. 

\item Argument registers that need to be preserved for use in the
first goal in the clause body.
\end{itemize}
Additional information may have to be recorded in this part depending 
on auxiliary language features. For example, in a framework that
permits the use of the {\it cut} control primitive, the contents of
the cut point register that indicates the backtracking point up to
which to eliminate choices needs to be stored as well. Similarly,
it has been found useful to give the programmer dynamic control over
whether projection or imitation substitutions are to be tried first
within higher-order unification. In this situation, the regimen in
effect at this instance should also be stored in the shared part for
later restoration. 

Once the shared part has been constructed, a global reference to
it is maintained in a special register that we refer to as the {\it
BRS} register. Computation now 
proceeds to simplifying the disagreement set and, eventually, to
picking a flexible-rigid pair whose imitation and projection
substitutions have to be examined. After such a pair has been determined
and a substitution for it has been selected, information must be left
behind for examining the remaining alternatives in case of 
backtracking. This data is encoded in the variable part of unification
branch point records that comprises the following components:
\begin{itemize}
\item The heads of the flexible and rigid terms together with their
types.

\item Information determining what substitutions remain to be tried. A
simple way to encode this is by remembering the number of projection
substitutions already tried; the additional knowledge of whether
imitation or projection substitutions are being tried first completely
determines the alternatives left. 

\item The contents of the {\it LL} register that will be used in
consort with the information in the trail stack to restore the
disagreement set on backtracking.\footnote{This component needs also
to be added to the usual {\it choice point record} of the WAM that
stores information for backtracking over clause choices.}

\item Pointers to the top of the heap and the trail stack that
determine the status of these data areas. 

\item The contents of the {\it BRS} register for restoring clause
context on backtracking.

\item A pointer to a record of the preceding branch point in
computation, to be used when all alternatives at this stage have been
exhausted. 
\end{itemize}
Although the heads of the flexible and rigid terms suffice for generating
all the substitutions, certain operations have to be repeated on these
in each case. Thus, the binder of every substitution that is
constructed is identical. Similarly, the vector of arguments of the
general arguments of the substitution terms are identical both within
a single substitution and across the imitation and
projections. Finally, the target 
type of the flexible head is used repeatedly in determining the
appropriateness of each projection substitution. Assuming the
acceptability of trading off space for time, these components may be
computed once when the variable part is set up and references to them
may be saved for later use. This is, in fact, the course adopted within the
{\it Teyjus} implementation.  

The successful selection of a substitution within the unification
process is followed by another term simplification phase. If there is
another flexible-rigid pair in the resulting disagreement set, further
substitutions must be posited, leading to the setting up of the
variable part of another unification branch point record. Note that
the shared part of this record, pointed to by the {\it BRS} register,
is the same as that for the previous such record. This process
continues till eventually a failure is encountered or the disagreement
set is reduced to a solved form. In the latter case, computation
continues with an attempt to solve the next (predicate) goal through a
backchaining process.

There is, of course, the possibility
of failure along the path currently being explored. In this case 
backtracking must take place to the most recent choice point either in
clause selection or in unification. In order to determine the
appropriate such point, the records corresponding to them are 
chained into one linear sequence based on their age and a pointer to
the most recent one is placed, as in the WAM, in the {\it B} or
backtrack register. Now, certain actions, such as the unwinding of the 
trail stack, the resetting of the disagreement set and recovery of
heap space, are identical regardless of whether computation returns
to trying another clause or another unifier and can be carried out
uniformly with a little coordination in the structures of the records
corresponding to these different kinds of backtrack points. However,
other actions, such as the generation of another substitution or the 
selection of another clause, do need a knowledge of the kind of
choice being reconsidered. One approach to providing this
information would be to mark each backtrack point record in a special
way at the time of its creation. A more elegant solution is
possible in a virtual machine and compilation based framework and is,
in fact, used in {\it Teyjus}. In this system, a special
instruction is included in the instruction set whose purpose is to
utilize the information in the variable part of a unification
branch point record to generate a new substitution, to reset the {\it
BRS} register and the state reflecting clause usage context and to
continue with the unification computation. The right backtracking
action can now be achieved simply by storing a pointer to a program
location containing this instruction in a field of the variable part
of a unification branch point record that is coordinated with the next 
clause field of the usual choice point record of the WAM; a uniform
transfer of control to the stored program point and the execution of
the corresponding instruction then achieves the appropriate
backtracking action.  

Branching in computation is obviously costly both in time and in
space and every effort should be expended to exploit deterministic
execution patterns whenever possible. One approach to doing this
within the unification computation is, as we have already mentioned,
to build a treatment of more special cases in which most general
unifiers exist into the term simplification process. Another useful
idea is to employ quick dynamic tests to determine that no further
substitutions exist in certain cases and to discard unification
backtrack points eagerly on this basis. Some heuristics of this kind
are embedded in the {\it Teyjus} system but this is a matter that
deserves further attention.

A final point to mention concerns the allocation of space for the
terms generated for projection and imitation substitutions. 
This is best done on the heap since backtracking permits the
space to be reclaimed when the substitution itself becomes redundant. 

\section{An Abstract Machine and Compilation Model}\label{sec:abstmachine}

The abstract machine for Prolog is designed to support a compiled
treatment of the four main components of the underlying model of
computation: the processing of the structure of complex, usually
conjunctive, goals, the setting up of the arguments of atomic goals,
the sequencing through clause choices for such goals and the
unification of the arguments of these goals with the statically known
arguments of a clause head.  A further aspect that receives special
attention is the detection of determinism. Nondeterminism is costly to
deal with and the need to do so can often be eliminated by utilizing
the structure of the actual arguments of atomic goals to prune choices
early during execution. This observation is exploited in practice by
including a special set of instructions that allow clause choices to
be indexed by the arguments and by building the use of these
instructions  into the compilation process.

The basic issues in a first-order context persist also in our
higher-order language. Much of the machinery and even the instruction
set that are embedded in a WAM-like architecture for treating these
aspects can, in fact, be carried over to the implementation task at
hand. However, some new devices are needed, primarily for dealing with
a richer structure for terms and a more complex unification
operation. Moreover, there must be differences in the interpretation
of some instructions. Instructions that examine the structure of terms
must, for instance, have the ability to head normalize these terms if
this is needed during execution. Further, the instructions that
realize unification completely in a first-order setting suffice only
to implement the initial term simplification phase of higher-order
unification. The capability to leave unification problems that cannot
be solved in this manner to a later, interpretive phase should
therefore be built into these instructions. Such a `deferring' action
should, of course, be complemented by an invocation of the remaining 
higher-order unification process at a suitably chosen point.

We present, in this section, an extended version of the WAM that
develops on these ideas. We summarize first the modifications to
the machine structure that were implicit in our discussion of
the treatment of higher-order unification. We then describe changes to
the instruction set. The last part of this section illustrates the
compilation model by presenting the code generated for some simple
higher-order programs. A familiarity with the original abstract
machine of Warren, such as might be obtained from \cite{HAK91} or
\cite{War83}, is assumed in this exposition.

\subsection{The Structure of the Extended Machine}\label{ssec:wam}

\begin{figure}
\begin{center}
\setlength{\unitlength}{0.00083333in}
\begingroup\makeatletter\ifx\SetFigFont\undefined%
\gdef\SetFigFont#1#2#3#4#5{%
  \reset@font\fontsize{#1}{#2pt}%
  \fontfamily{#3}\fontseries{#4}\fontshape{#5}%
  \selectfont}%
\fi\endgroup%
{\renewcommand{\dashlinestretch}{30}
\begin{picture}(5877,8787)(0,-10)
\path(12,5862)(1512,5862)
\path(762,5862)(762,5487)
\blacken\path(732.000,5607.000)(762.000,5487.000)(792.000,5607.000)(762.000,5571.000)(732.000,5607.000)
\path(1887,6612)(1512,6612)
\blacken\path(1632.000,6642.000)(1512.000,6612.000)(1632.000,6582.000)(1596.000,6612.000)(1632.000,6642.000)
\path(12,5562)(12,6837)(1512,6837)(1512,5562)
\path(1887,6237)(1512,6237)
\blacken\path(1632.000,6267.000)(1512.000,6237.000)(1632.000,6207.000)(1596.000,6237.000)(1632.000,6267.000)
\path(1887,5862)(1512,5862)
\blacken\path(1632.000,5892.000)(1512.000,5862.000)(1632.000,5832.000)(1596.000,5862.000)(1632.000,5892.000)
\put(2037,6537){\makebox(0,0)[lb]{\smash{{{\SetFigFont{10}{12.0}{\rmdefault}{\mddefault}{\updefault}LL}}}}}
\put(2037,6162){\makebox(0,0)[lb]{\smash{{{\SetFigFont{10}{12.0}{\rmdefault}{\mddefault}{\updefault}HB}}}}}
\put(2037,5787){\makebox(0,0)[lb]{\smash{{{\SetFigFont{10}{12.0}{\rmdefault}{\mddefault}{\updefault}H}}}}}
\put(462,6987){\makebox(0,0)[lb]{\smash{{{\SetFigFont{12}{14.4}{\rmdefault}{\bfdefault}{\updefault}Heap}}}}}
\put(162,312){\makebox(0,0)[lb]{\smash{{{\SetFigFont{12}{14.4}{\rmdefault}{\bfdefault}{\updefault}Applications}}}}}
\put(162,87){\makebox(0,0)[lb]{\smash{{{\SetFigFont{12}{14.4}{\rmdefault}{\bfdefault}{\updefault}Stack}}}}}
\path(12,612)(12,2937)(1512,2937)
	(1512,612)(12,612)
\path(12,2187)(1512,2187)
\path(12,1212)(1512,1212)
\path(762,2187)(762,1812)
\blacken\path(732.000,1932.000)(762.000,1812.000)(792.000,1932.000)(762.000,1896.000)(732.000,1932.000)
\path(762,1212)(762,1587)
\blacken\path(792.000,1467.000)(762.000,1587.000)(732.000,1467.000)(762.000,1503.000)(792.000,1467.000)
\path(1887,2487)(1512,2487)
\blacken\path(1632.000,2517.000)(1512.000,2487.000)(1632.000,2457.000)(1596.000,2487.000)(1632.000,2517.000)
\path(1887,2187)(1512,2187)
\blacken\path(1632.000,2217.000)(1512.000,2187.000)(1632.000,2157.000)(1596.000,2187.000)(1632.000,2217.000)
\put(2037,2412){\makebox(0,0)[lb]{\smash{{{\SetFigFont{10}{12.0}{\rmdefault}{\mddefault}{\updefault}S}}}}}
\put(2037,2112){\makebox(0,0)[lb]{\smash{{{\SetFigFont{10}{12.0}{\rmdefault}{\mddefault}{\updefault}SL}}}}}
\put(312,3087){\makebox(0,0)[lb]{\smash{{{\SetFigFont{12}{14.4}{\rmdefault}{\bfdefault}{\updefault}SL Stack}}}}}
\path(12,3762)(12,4887)(1512,4887)(1512,3762)
\path(12,4062)(1512,4062)
\path(762,4062)(762,3687)
\blacken\path(732.000,3807.000)(762.000,3687.000)(792.000,3807.000)(762.000,3771.000)(732.000,3807.000)
\path(1887,4062)(1512,4062)
\blacken\path(1632.000,4092.000)(1512.000,4062.000)(1632.000,4032.000)(1596.000,4062.000)(1632.000,4092.000)
\put(2037,3987){\makebox(0,0)[lb]{\smash{{{\SetFigFont{10}{12.0}{\rmdefault}{\mddefault}{\updefault}TR}}}}}
\put(537,5037){\makebox(0,0)[lb]{\smash{{{\SetFigFont{12}{14.4}{\rmdefault}{\bfdefault}{\updefault}Trail}}}}}
\path(3012,12)(3012,3237)(4812,3237)(4812,12)
\path(3012,2862)(4812,2862)
\path(3012,2562)(4812,2562)
\path(3012,2187)(4812,2187)
\path(5187,1887)(4812,1887)
\blacken\path(4932.000,1917.000)(4812.000,1887.000)(4932.000,1857.000)(4896.000,1887.000)(4932.000,1917.000)
\path(3012,1887)(4812,1887)
\dashline{60.000}(4812,687)(5112,687)(5112,912)(4812,912)
\blacken\path(4932.000,942.000)(4812.000,912.000)(4932.000,882.000)(4896.000,912.000)(4932.000,942.000)
\dashline{60.000}(4812,987)(5112,987)(5112,1212)(4812,1212)
\blacken\path(4932.000,1242.000)(4812.000,1212.000)(4932.000,1182.000)(4896.000,1212.000)(4932.000,1242.000)
\path(3012,1737)(4812,1737)
\path(3012,1137)(4812,1137)
\path(3012,837)(4812,837)
\path(3012,612)(2487,612)(2487,1662)(3012,1662)
\blacken\path(2892.000,1632.000)(3012.000,1662.000)(2892.000,1692.000)(2928.000,1662.000)(2892.000,1632.000)
\path(3012,912)(2637,912)(2637,1587)(3012,1587)
\blacken\path(2892.000,1557.000)(3012.000,1587.000)(2892.000,1617.000)(2928.000,1587.000)(2892.000,1557.000)
\path(3012,1212)(2787,1212)(2787,1512)(3012,1512)
\blacken\path(2892.000,1482.000)(3012.000,1512.000)(2892.000,1542.000)(2928.000,1512.000)(2892.000,1482.000)
\dashline{60.000}(4812,1287)(5862,1287)(5862,2712)(4812,2712)
\blacken\path(4932.000,2742.000)(4812.000,2712.000)(4932.000,2682.000)(4896.000,2712.000)(4932.000,2742.000)
\path(5187,1587)(4812,1587)
\blacken\path(4932.000,1617.000)(4812.000,1587.000)(4932.000,1557.000)(4896.000,1587.000)(4932.000,1617.000)
\path(3012,537)(4812,537)
\path(2937,1437)(4737,1437)
\path(3837,537)(3837,162)
\blacken\path(3807.000,282.000)(3837.000,162.000)(3867.000,282.000)(3837.000,246.000)(3807.000,282.000)
\put(3387,2637){\makebox(0,0)[lb]{\smash{{{\SetFigFont{10}{12.0}{\rmdefault}{\mddefault}{\updefault}Choice Point}}}}}
\put(3087,1962){\makebox(0,0)[lb]{\smash{{{\SetFigFont{10}{12.0}{\rmdefault}{\mddefault}{\updefault}Environment Record}}}}}
\put(5262,1812){\makebox(0,0)[lb]{\smash{{{\SetFigFont{10}{12.0}{\rmdefault}{\mddefault}{\updefault}E}}}}}
\put(5262,1512){\makebox(0,0)[lb]{\smash{{{\SetFigFont{10}{12.0}{\rmdefault}{\mddefault}{\updefault}BRS}}}}}
\put(5262,537){\makebox(0,0)[lb]{\smash{{{\SetFigFont{10}{12.0}{\rmdefault}{\mddefault}{\updefault}B}}}}}
\put(3087,912){\makebox(0,0)[lb]{\smash{{{\SetFigFont{10}{12.0}{\rmdefault}{\mddefault}{\updefault}Branch Pt. (Variable)}}}}}
\put(3237,3387){\makebox(0,0)[lb]{\smash{{{\SetFigFont{12}{14.4}{\rmdefault}{\bfdefault}{\updefault}Local Stack}}}}}
\path(5187,612)(4812,612)
\blacken\path(4932.000,642.000)(4812.000,612.000)(4932.000,582.000)(4896.000,612.000)(4932.000,642.000)
\drawline(3162,1212)(3162,1212)
\put(3087,1212){\makebox(0,0)[lb]{\smash{{{\SetFigFont{10}{12.0}{\rmdefault}{\mddefault}{\updefault}Branch Pt. (Variable)}}}}}
\put(3087,612){\makebox(0,0)[lb]{\smash{{{\SetFigFont{10}{12.0}{\rmdefault}{\mddefault}{\updefault}Branch Pt. (Variable)}}}}}
\put(3012,1512){\makebox(0,0)[lb]{\smash{{{\SetFigFont{10}{12.0}{\rmdefault}{\mddefault}{\updefault}Branch Point (Shared)}}}}}
\path(4737,8337)(4738,8338)(4744,8344)
	(4756,8355)(4770,8368)(4783,8379)
	(4794,8388)(4803,8395)(4812,8399)
	(4819,8403)(4826,8405)(4834,8407)
	(4841,8408)(4850,8409)(4858,8408)
	(4865,8407)(4873,8405)(4880,8403)
	(4887,8399)(4894,8396)(4901,8391)
	(4908,8385)(4916,8378)(4923,8369)
	(4930,8360)(4936,8349)(4941,8337)
	(4946,8325)(4950,8312)(4952,8300)
	(4954,8287)(4957,8273)(4958,8257)
	(4960,8241)(4962,8224)(4964,8208)
	(4966,8192)(4967,8176)(4970,8162)
	(4972,8149)(4975,8137)(4978,8124)
	(4983,8111)(4988,8100)(4994,8088)
	(5000,8078)(5005,8068)(5011,8059)
	(5016,8051)(5021,8044)(5025,8037)
	(5028,8030)(5030,8023)(5032,8015)
	(5033,8008)(5034,7999)(5033,7991)
	(5032,7984)(5030,7976)(5028,7969)
	(5025,7962)(5021,7955)(5016,7948)
	(5011,7940)(5005,7931)(4999,7921)
	(4994,7911)(4988,7899)(4983,7888)
	(4978,7875)(4975,7862)(4972,7850)
	(4970,7837)(4967,7823)(4966,7807)
	(4964,7791)(4962,7774)(4960,7758)
	(4958,7742)(4957,7726)(4954,7712)
	(4952,7699)(4950,7687)(4946,7674)
	(4941,7662)(4936,7650)(4930,7639)
	(4923,7630)(4916,7621)(4908,7614)
	(4901,7608)(4894,7603)(4887,7599)
	(4880,7596)(4873,7594)(4865,7592)
	(4858,7591)(4849,7590)(4841,7591)
	(4834,7592)(4826,7594)(4819,7596)
	(4812,7599)(4803,7604)(4794,7611)
	(4783,7620)(4770,7631)(4756,7644)
	(4744,7655)(4738,7661)(4737,7662)
\put(3087,7737){\makebox(0,0)[lb]{\smash{{{\SetFigFont{10}{12.0}{\rmdefault}{\mddefault}{\updefault}S, HB, A1, ..., An}}}}}
\put(3087,8037){\makebox(0,0)[lb]{\smash{{{\SetFigFont{10}{12.0}{\rmdefault}{\mddefault}{\updefault}P, CP, E, H, B, TR,}}}}}
\put(5262,8112){\makebox(0,0)[lb]{\smash{{{\SetFigFont{10}{12.0}{\rmdefault}{\mddefault}{\updefault}From the }}}}}
\put(5262,7887){\makebox(0,0)[lb]{\smash{{{\SetFigFont{10}{12.0}{\rmdefault}{\mddefault}{\updefault}WAM}}}}}
\path(4737,7362)(4737,7363)(4739,7366)
	(4743,7373)(4748,7383)(4756,7395)
	(4764,7407)(4774,7418)(4785,7428)
	(4797,7434)(4812,7437)(4822,7436)
	(4832,7433)(4839,7430)(4845,7426)
	(4850,7423)(4853,7421)(4856,7418)
	(4858,7415)(4861,7412)(4865,7406)
	(4870,7399)(4875,7390)(4881,7377)
	(4887,7362)(4892,7343)(4895,7326)
	(4895,7312)(4893,7301)(4890,7293)
	(4887,7287)(4884,7281)(4881,7273)
	(4879,7262)(4879,7248)(4882,7231)
	(4887,7212)(4896,7192)(4905,7178)
	(4914,7171)(4923,7168)(4931,7168)
	(4939,7168)(4946,7167)(4954,7161)
	(4959,7151)(4962,7137)(4959,7123)
	(4954,7113)(4946,7107)(4939,7106)
	(4931,7106)(4923,7106)(4914,7103)
	(4905,7096)(4896,7082)(4887,7062)
	(4882,7043)(4879,7026)(4879,7012)
	(4881,7001)(4884,6993)(4887,6987)
	(4890,6981)(4893,6973)(4895,6962)
	(4895,6948)(4892,6931)(4887,6912)
	(4881,6897)(4875,6884)(4870,6875)
	(4865,6868)(4861,6862)(4858,6859)
	(4856,6856)(4853,6853)(4850,6851)
	(4845,6848)(4839,6844)(4832,6841)
	(4822,6838)(4812,6837)(4797,6840)
	(4785,6846)(4774,6856)(4764,6867)
	(4756,6879)(4748,6891)(4743,6901)
	(4739,6908)(4737,6911)(4737,6912)
\put(3087,7212){\makebox(0,0)[lb]{\smash{{{\SetFigFont{10}{12.0}{\rmdefault}{\mddefault}{\updefault}LL, SL, BRS,}}}}}
\put(3087,6912){\makebox(0,0)[lb]{\smash{{{\SetFigFont{10}{12.0}{\rmdefault}{\mddefault}{\updefault}NUMARGS}}}}}
\put(5262,7137){\makebox(0,0)[lb]{\smash{{{\SetFigFont{10}{12.0}{\rmdefault}{\mddefault}{\updefault}New}}}}}
\put(3312,8412){\makebox(0,0)[lb]{\smash{{{\SetFigFont{12}{14.4}{\rmdefault}{\bfdefault}{\updefault}Registers}}}}}
\path(12,8487)(1512,8487)(1512,7437)
	(12,7437)(12,8487)
\path(1887,8187)(1512,8187)
\blacken\path(1632.000,8217.000)(1512.000,8187.000)(1632.000,8157.000)(1596.000,8187.000)(1632.000,8217.000)
\path(1887,7812)(1512,7812)
\blacken\path(1632.000,7842.000)(1512.000,7812.000)(1632.000,7782.000)(1596.000,7812.000)(1632.000,7842.000)
\path(3912,4662)(3912,5037)
\blacken\path(3942.000,4917.000)(3912.000,5037.000)(3882.000,4917.000)(3912.000,4953.000)(3942.000,4917.000)
\path(3912,5787)(3912,5412)
\blacken\path(3882.000,5532.000)(3912.000,5412.000)(3942.000,5532.000)(3912.000,5496.000)(3882.000,5532.000)
\put(237,8637){\makebox(0,0)[lb]{\smash{{{\SetFigFont{12}{14.4}{\rmdefault}{\bfdefault}{\updefault}Code Area}}}}}
\put(2037,8112){\makebox(0,0)[lb]{\smash{{{\SetFigFont{10}{12.0}{\rmdefault}{\mddefault}{\updefault}P}}}}}
\put(2037,7737){\makebox(0,0)[lb]{\smash{{{\SetFigFont{10}{12.0}{\rmdefault}{\mddefault}{\updefault}CP}}}}}
\put(3387,3987){\makebox(0,0)[lb]{\smash{{{\SetFigFont{12}{14.4}{\rmdefault}{\bfdefault}{\updefault}Second PDL}}}}}
\put(3462,6237){\makebox(0,0)[lb]{\smash{{{\SetFigFont{12}{14.4}{\rmdefault}{\bfdefault}{\updefault}First PDL}}}}}
\path(3387,5037)(3387,4212)(4512,4212)(4512,5037)
\path(3387,4662)(4512,4662)
\path(3387,5337)(3387,6162)(4512,6162)(4512,5337)
\path(3387,5787)(4512,5787)
\end{picture}
}

\end{center}
\caption{Abstract Machine Data Areas and Snapshot of State}
\label{fig:wamstate}
\end{figure}
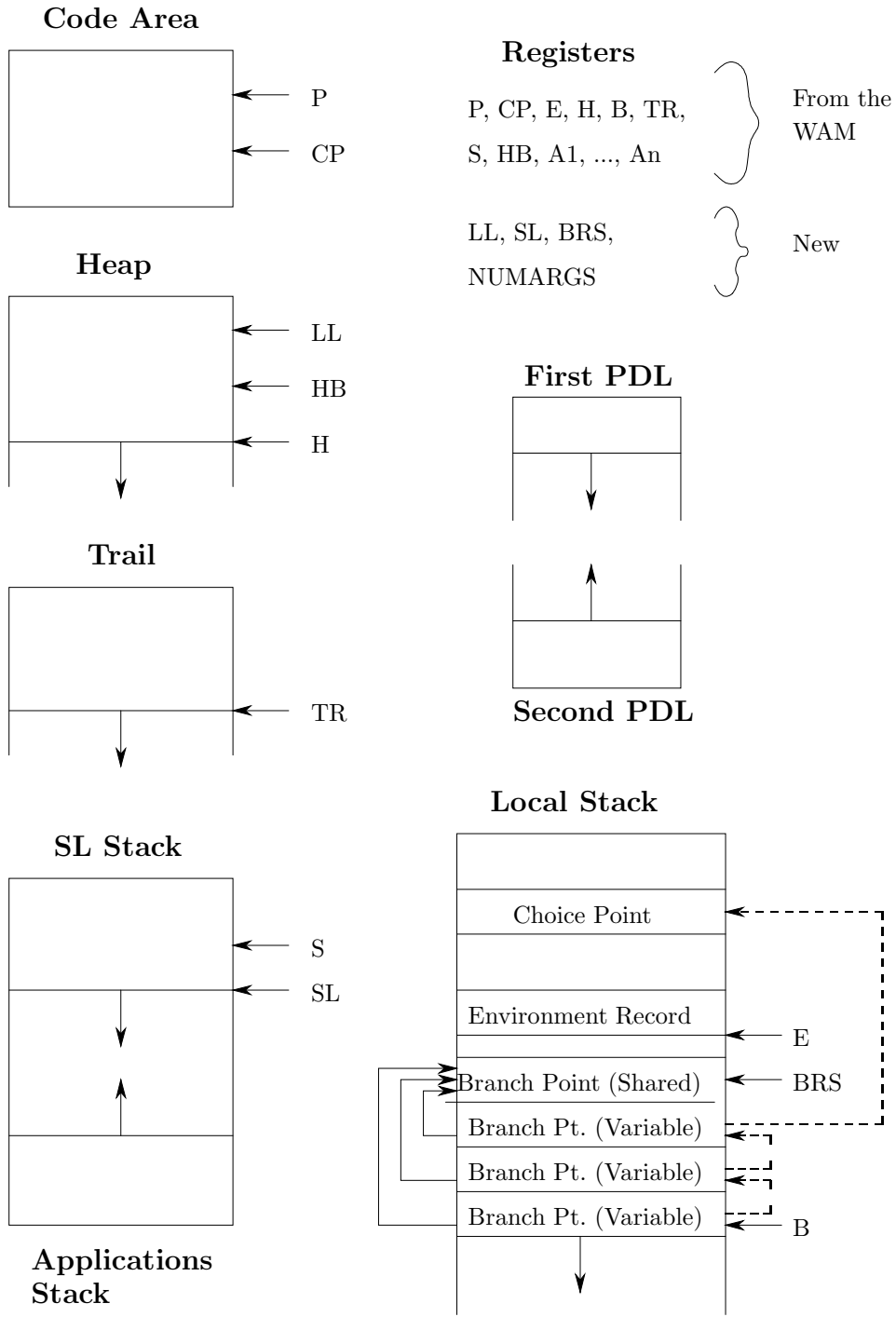

Figure~\ref{fig:wamstate} depicts the various data areas and registers
present in the extended abstract machine and provides a snapshot of a
machine state during computation. The code area, the heap, the local
stack and the trail of the WAM persist in this machine. The new data
areas are the {\it SL} stack, the applications stack and the two
pushdown lists. The first two components are utilized by the
head normalization code as described in Section~\ref{sec:runtime}. The
pushdown lists are used in simplifying disagreement sets and help, as
we have seen, in conserving heap space. We observe that only
one of these pushdown lists is really new: one pushdown list is
usually employed by WAM implementations for realizing the part of
first-order unification that must be performed in interpretive
mode.

While several data areas are carried over from the WAM, their usage in
our machine differs in certain respects. In addition to storing
compound terms that are created in a structure copying implementation,
the heap is used in our context also to store disagreement pairs, the
new terms that are generated during term reduction and the projection
and imitation substitutions generated by higher-order
unification. Similarly, the trail records not only the substitutions
made for variables, but also the destructive changes made to terms
during normalization and pointers to the pairs of terms removed from
disagreement sets in the course of term simplification. Of particular
note in this context are the facts that the trailing of terms requires
also that old values be stored and that different kind of entries
entail different unwinding actions and must be annotated appropriately
for determining this. Finally, in addition to the usual choice point
and environment records, the local stack must also store information
about branch points in unification. These are distinguished by being
labelled as {\it branch points} in Figure~\ref{fig:wamstate} that also
depicts their split representation between a shared and a variable
part. Only the variable parts of these records participate directly in
the chain of backtracking records; the shared parts gain currency by
being used by the variable parts. In the figure, we have used solid
arrows to depict the shared part and the variable part of a branch
point record and dashed arrows to depict the chain of branch points
and choice points that determine backtracking behaviour. 

The extended machine also includes a few new registers: the {\it LL}
register indicating the currently active disagreement set, the {\it
SL} register indicating the current top of the {\it SL} stack, the
{\it BRS} register indicating the currently relevant shared part of
branch point records and the {\it NUMARGS} register that holds the
(current) arity of an application encountered during head
normalization.  One slightly intriguing aspect of our depiction 
of the machine state is the fact that the {\it S} register that
indicates the argument vector of a compound term during unification is
shown pointing into the {\it SL} stack rather than the heap. The
reason for this is that the top-level, head normalized structure of a
higher-order term may become apparent only after a reduction process
and, in this case, is available as a vector only in the {\it SL}
stack. In special cases, such as when dealing with first-order terms,
no reduction steps are necessary and our representation of such terms
stores the arguments as a vector. Such situations can be recognized
and, as an optimization, the {\it S} register can be made to point to
the vector that is already available in the heap instead.

\subsection{Modifications to the Instruction Set}\label{ssec:instset}

A compilation model for our language must account for certain new
aspects in comparison with the one for Prolog. These aspects include a
representation of terms that differs even over the first-order
fragment, the possibility for function variables and abstractions to
appear in terms, the need to realize higher-order unification and the
necessity to treat mixed intensional and extensional uses of predicate
terms. We discuss these aspects in more detail below and we also
indicate changes to the instruction set of the WAM that are geared
towards treating them.

\medskip

\noindent {\bf Creating Typed, Higher-Order Terms.} The usual
compilation model requires that the arguments of atomic goals
appearing in the bodies of clauses be set up in registers prior to the
invocation of code for the relevant procedures. In the case that such
an argument is a compound term, its representation must be created in
the heap with a reference to it being placed in the relevant
register. These effects are actually realized through the {\it put}
and {\it unify} classes of instructions present in the WAM,
the latter being executed in write mode in these particular
situations. 

This basic structure carries over well to our higher-order
language and many specific instructions from the WAM can even be
retained for processing first-order like structure. There are,
however, two exceptions. First, in our context, types are retained with
variables and the instructions that create them must, for this reason,
take an extra type argument. In particular, these instructions might
take on the forms
\begin{tabbing}
\qquad\=\kill
\>{\it put\_variable Vi,Aj,type}, and \\
\>{\it unify\_variable Vi,type}
\end{tabbing}
\noindent where {\it Vi} is either a permanent or temporary variable
and {\it type} is a reference to the representation of a
type. The second difference arises from the modified representation of
a structure. We encode this as an application whose argument part is a
pointer to a vector with a size matching the arity of the
application. Moreover, in the general case, the 
`function' part of the application could be different from a
constant. In light of this, the {\it put\_structure} instruction might
be generalized to a {\it put\_app} instruction that fashions an
application on the heap. The abstract machine underlying the {\it
Teyjus} system, in fact, includes two instructions of the form
\begin{tabbing}
\qquad\=\kill
\>{\it put\_capp Ai,Xj,n}, and \\
\>{\it put\_fapp Ai,Xj,n}
\end{tabbing}
\noindent for this purpose. Each of these instructions creates an
application whose function part is obtained from the register {\it
Xj} and leaves a reference to this application in the register {\it
Ai}. Moreover, the application that is created has arity {\it n}, a
fact that is realized by allocating a vector of this size in the heap
for the argument part and by preparing to fill in these arguments by
setting the {\it S} register to the beginning of this vector and
turning the write mode on. The difference between the two instructions
is that the first annotates the application as closed whereas the
second annotates it as (possibly) open. 

The higher-order nature of our terms can manifest itself in three
ways in the syntax: the function part may be a variable, abstraction
may be explicitly present and there may be occurrences of abstracted
variables. The instruction for creating applications already accounts
for the special case of a variable `functor.' To support the creation
of abstractions, new instructions may be added to the {\it put} and
{\it unify} classes. The abstract machine for {\it Teyjus}
includes the following instructions for this purpose:
\begin{tabbing}
\qquad\={\it put\_clambda Ai,Xj}\qquad\qquad\=\kill
\>{\it put\_clambda Ai,Xj}\>{\it unify\_clambda Xj}\\
\>{\it put\_flambda Ai,Xj}\>{\it unify\_flambda Xj}
\end{tabbing}
\noindent The {\it put} versions create abstractions whose bodies are
given by the contents of register {\it Xj} on the heap and put
references to these abstractions in the register {\it
Ai}.\footnote{The register {\it Xj} may contain a constant, in which
case the first action of these instructions is to convert {\it Xj}
into a reference to a location on heap containing this constant.} The
difference between the two instructions provided for this purpose is
that one creates closed abstractions and the other open
ones. The {\it unify} versions, that are only ever executed in write
mode,  create similar abstractions but eventually put references to
these in the heap location pointed to by the {\it S} register and also
increment this register at the end. Finally, to support the creation
of bound variables, represented using indices in the de Bruijn scheme,
the following instructions in which {\it n} is a positive number, are
included in our abstract machine: 
\begin{tabbing}
\qquad\={\it put\_index Ai,n}\qquad\qquad\=\kill
\>{\it put\_index Ai,n}\>{\it unify\_index n}
\end{tabbing}
The first instruction writes a bound variable with index {\it n} on
the heap and makes the register {\it Ai} a reference to this
location. The {\it unify\_index} instruction, which, also is only
executed in write mode, stores this bound variable in the location
pointed to by the {\it S} register and then increments this register.

In the instructions that create applications and abstractions, the
function part and the abstraction body are both obtained from
registers. However, these components may in particular situations 
correspond to permanent variables. Furthermore, they may actually
dereference to stack cells that must be globalized prior to use. In
light of these possibilities, our abstract machine includes the
instructions {\it globalize Yi,Xj} and {\it globalize Xj}. The first
instruction dereferences the permanent (environment) variable {\it
Yi}. If this turns out to be a reference to the stack, then the value
is copied to the heap and the stack cell and the register {\it Xj} are
both converted into references to the newly created heap
cell. Otherwise the reference that we get to a heap cell is also
stored in the register {\it Xj}. The second instruction simply
dereferences the {\it Xj} register, globalizes this as before if
necessary and leaves a reference to a suitable heap cell in {\it Xj}.

\medskip

\noindent {\bf Compilation of Higher-Order Unification.} In any given
use of a clause, the terms that appear as arguments of the head of a
clause must be unified with the terms that arrive in the relevant
argument registers. The compilation model for Prolog translates each
of these statically known terms into a sequence of instructions that
either creates a relevant term that the incoming argument is bound to
if this argument is an uninstantiated variable and that carries out an
analysis of the structure of the argument if it is not a
variable. This model requires the same instructions to function in two 
different dynamically determined ways, an ability that is realized
through the use of the read and write modes. 

Lifting this treatment of unification to cover the operation in its
entirety in the higher-order situation is difficult. In particular,
statically available structure is not directly usable once a function
variable with arguments is reached in it and is also difficult to
exploit when a flexible, nonvariable part is exposed in the incoming
term relative to which a set of matching substitutions have to be
tried. However, at least the first phase of term simplification can be
compiled and, if augmented with the simple forms of variable bindings
discussed in Section~\ref{ssec:disset}, most of the unification
computation that arises in practice can be treated completely within
this phase. 

The {\it get} and {\it unify} class of WAM instructions that treat
head unification can, in fact, be adapted to realize this idea when 
the term to be compiled has a first-order structure at the top level,
\ie, when it is a variable, a constant or an application in which the
head is a constant.\footnote{Since this term will be normalized prior
to compilation, the only remaining possibilities are that it is an
abstraction or an application with a variable at the head.} However, a
few changes in interpretation are necessary for the instructions {\it
get\_structure}, {\it get\_constant} and {\it unify\_constant} that
are used in compiling rigid structure. First, these instructions must
take responsibility for head normalizing the input term at the
outset. In practice, many of these terms have a first-order structure,
a fact that can be recognized through a check of their heads (that
constitutes an overhead only when the terms are {\it not} first-order
ones) built into 
the relevant instructions so that an explicit invocation of
head normalization can be avoided. Notice that the {\it
get\_structure} instruction must set the {\it S} register to point to
the vector of arguments in case the incoming term is itself an
application with the right head and, under the considered
optimization, this would become a pointer either into the {\it SL}
stack or into the heap.  The second change is that when the incoming
term is a variable, the {\it get\_structure} instruction must create
an application of a specified arity on the heap and so should get this
arity as an additional argument. In the {\it Teyjus} abstract machine,
the instruction actually has the format
\begin{tabbing}
\qquad\=\kill
\>{\it get\_structure Ai,f,n}
\end{tabbing}
\noindent where {\it n} is a positive number; when executed in a mode
in which a term has to be created, this instruction pushes an
application with arity {\it n} and function part {\it f} onto the
heap, followed by a vector of size {\it n} constituting the argument
part of this application and sets the {\it S} register to the
beginning of this vector.  The final change arises from the fact that
these instructions must also cater to the possibility that the
incoming term is an application with a flexible head. One possible
strategy in such a case is to add a suitable pair to the existing
disagreement set and to leave its further processing to a later
interpretive treatment of genuine higher-order unification. In the
situation where the instruction is {\it get\_structure}, we note that
the added disagreement pair will actually involve a term that is
created by subsequent actions carried out by this and following
instructions executed in write mode.

It is in principle possible to extend the compilation of first-order
like structure to include the case of terms that have abstractions at
their head. However, it is not clear if enough situations where this
is needed will occur in practice so as to make such a treatment
pragmatically useful. We therefore describe a simpler approach that
works uniformly for this case as well as for the last remaining case
which is that of an application whose head is a variable. In essence,
the term in both situations may be translated into a sequence of
instructions that constructs its representation and leaves a reference
to it in a register, followed by an instruction that invokes term
simplification in interpretive mode. We have already discussed
instructions for creating higher-order terms. To realize the last
effect, we may use the {\it get\_value} instruction from the WAM that,
in any case, has to be adapted to deal with higher-order terms. In
particular, in the new form, the instruction invokes an interpretive
phase of term simplification that may make simple bindings for
variables and that may add new flexible-rigid pairs to the existing
disagreement set. A similar kind of generalization
must be made to the {\it unify\_value} instruction. Actually, another
change to these instructions is also necessary. Although usual
implementations of Prolog omit occurs-checks, the place to carry these
out if they are included would be within the process invoked by the
{\it get\_value} and the {\it unify\_value} instructions. The situation
in the higher-order case is similar, except that rigid path checks
would replace occurs checks. These checks turn out to be indispensable
to the envisaged applications of the language whose implementation we
are considering, and so they are included in the `higher-order'
versions of the {\it get\_value} and {\it unify\_value}
instructions.\footnote{These rigid path checks need the complete
  normalization of incoming terms, bringing up an interesting
  question: is there still an advantage to laziness in substitution
  and reduction? This issue is examined in detail in \cite{LNX03}. The 
  conclusion from this study is, briefly, that a demand driven
  approach to reduction that exploits explicit substitutions has
  significant advantages even if the particular style chosen in
  this paper is not uniquely the best.}
Now, as discussed in Section~\ref{ssec:disset}, a rigid path check may
permit only a partial instantiation of a variable, the rest of the
instantiation being subject to the resolution of constraints
represented by new flexible terms and relevant subparts of the
incoming terms. 
When creating these terms, the type of the new (logic) variables at
their heads must be written to the heap. These types can be generated
from a knowledge of the type of the variable whose compilation yields
the {\it unify\_value} instruction and, conversely, the type of this
variable is needed at least in the case when the constraint involves
the entire incoming term. 
In keeping with this observation, the {\it unify\_value} instruction 
acquires a type as an additional argument.

After simplification has been carried out relative to all the terms
appearing in the head of a clause, it may be necessary to invoke an
interpretive phase of higher-order unification. Our abstract machine
includes three instructions for this purpose. One of these, the {\it
proceed\_finish\_unify} instruction, is used in place of the {\it
proceed} instruction of the WAM in the situation when the clause body
is empty and when an unresolved higher-order unification problem may
exist. The effect of this instruction is to set the program pointer to
the continuation point, to set up the shared part of a branch point
record and, finally, to invoke code that tries to complete the
unification process. The code that is invoked tries to generate a
matching substitution. If one is found, then this is applied to the
state, the variable part of a branch point record representing the
remaining matching substitutions is created and the simplification
and substitution generation processes are iterated.  A point to note
about the situation in which {\it proceed\_finish\_unify} is used is
that no argument registers need to be stored in the shared part of the
branch point record. The second instruction, {\it
execute\_finish\_unify}, is used when the body of the clause consists
of a single atomic goal. This instruction differs from {\it
proceed\_finish\_unify} in that it must update the program pointer to
the next instruction in sequence and also save the continuation point
and relevant argument registers in the shared part of the branch point
record for use on backtracking. The number of argument registers
that must be remembered becomes a parameter to this instruction. The
final instruction, {\it call\_finish\_unify}, is used when the body of
the clause has multiple goals in it, and therefore requires an
environment record to be created for its invocation. This instruction
behaves differently from {\it execute\_finish\_unify} in only two
respects. First, it does not need to save the continuation point since
this is available from the environment record. Second, before
it allocates space for the shared part of the branch point record, the
instruction must ensure that sufficient space has been left for the
permanent variables in the clause. On account of the latter
requirement, {\it call\_finish\_unify} acquires the count of 
the permanent variables as an argument, in addition to the count of
the register arguments that need to be saved.

The interpretive phase of unification is, of course, not always
needed. In particular, it need only be considered if the compiled
form of term simplification leads to additions to the original
disagreement set or to bindings for variables that have the
potential of modifying the status of existing pairs. Compile-time
analysis can sometimes determine this cannot happen and,
consequently, that the new instructions need not be used. Even when the
instructions are included in the compiled code, they can incorporate a
checking of flags set during the compiled term simplification phase to
determine if further processing is necessary. The {\it Teyjus}
implementation utilizes such ideas to avoid unnecessary examination of
disagreement sets and setting up of the shared parts of branch point
records. 

\medskip

\noindent {\bf Treating Mixed Uses of Predicates.} The crux of this
treatment is the compilation of flexible atomic goals: mixed uses of
predicate terms arises essentially from the fact that flexible goals
may be instantiated by terms with complex logical structure, thereby
reflecting intensional occurrences of quantifiers and connectives into
positions where they function as search directives. 

The problem in the treatment of flexible atomic goals is, of course,
that their top-level structure is determined dynamically, and so the
specific action to be performed is not known at compilation time.
Nevertheless, some part of the action can be compiled by using the
knowledge of the possible cases that can arise. In particular,
flexible goals can be compiled into calls to a special procedure named 
{\it solve} to which (the instantiated version of) the goal is
provided as an argument. In the case that (the normalized form of) the
instantiated goal has a complex structure, the behavior of {\it solve}
can be envisaged as if it were based on a compilation of the following
clauses in which we use semicolon to represent disjunction in an
extensional position:
\begin{tabbing}
\qquad\=\kill
\>${\it solve \app (G1 \land G2)}$ {\tt :-} ${\it (solve\app G1),
(solve\app G2)}.$\\ 
\>${\it solve\app (G1 \lor G2)}$ {\tt :-} ${\it (solve\app G1) ; (solve\app G2)}.$\\
\>${\it solve\app (\Sigma \app G)}$ {\tt :-} ${\it solve\app (G\app X)}.$
\end{tabbing}
\noindent To complete the description of {\it solve}, it only remains
to specify its behaviour in the situation when its argument is an atomic
goal. In the case that this goal is a flexible one, {\it solve}
succeeds after instantiating the head of the goal to a term of the
form $\lambda\, \ldots \lambda\, \top$, the binder being chosen based
on type considerations. If this goal is a rigid one, then its
arguments are loaded into appropriate argument registers and the head
is used to determine the code to be invoked next. 

In the {\it Teyjus} implementation, the {\it solve} predicate is
treated as a builtin one whose realization is `hard-wired' into 
that of the abstract machine. 

\subsection{Examples of Compiled Code}

Based on the compilation scheme described in this section, code of the
following form might be generated from the definition of the ${\it
mapfun}$ predicate presented in Section~\ref{sec:hohc}:
\begin{tabbing}
\qquad\={\it mapfun:}\quad\={\it switch\_on\_term L2, L3, L5, fail}\quad\=\kill
\>{\it mapfun:}\>{\it switch\_on\_term L2, L3, L5, fail}\>{\it \%}\\
\>{\it L2:}\>{\it try\_me\_else L4, 3}\>{\it \% mapfun}\\
\>{\it L3:}\>{\it get\_nil A1}\>{\it \% nil}\\
\>\>{\it get\_nil A3}\> {\it \% F nil}\\
\>\>{\it proceed\_finish\_unify}\>{\it \%}\\
\>{\it L4:}\>{\it trust\_me 3}\>{\it \% mapfun}\\
\>{\it L5:}\>{\it get\_list A1}\>{\it \% (::} \\
\>\>{\it unify\_variable A4, ty1}\> {\it \% X}\\
\>\>{\it unify\_variable A1, ty2}\> {\it \%  L1)}\\
\>\>{\it get\_list A3}\>{\it \% F (:: }\\
\>\>{\it unify\_variable A5, ty1}\>{\it \% S1}\\
\>\>{\it unify\_variable A3, ty2}\> {\it \% L2)}\\
\>\>{\it globalize A2}\>{\it \%}\\
\>\>{\it put\_capp A6, A2, 1}\> {\it \% S2 = (F }\\
\>\>{\it unify\_value A4, ty1}\> {\it \% X)}\\
\>\>{\it get\_value A6, A5}\>{\it \% S1 = S2}\\
\>\>{\it execute\_finish\_unify 3}\>{\it \% :-}\\
\>\>{\it execute mapfun}\>{\it \% mapfun L1 F L2}
\end{tabbing}
\noindent This code uses the instructions {\it get\_nil} and {\it
get\_list} that realize, as in the WAM, special cases of the {\it
get\_constant} and {\it get\_structure} instructions. Also used is
the instruction {\it switch\_on\_term} that adapts an indexing
instruction with the same name from the WAM. In our context, this
instruction takes the form 
\begin{tabbing}
\qquad\=\kill
\> {\it switch\_on\_term V,C,L,BV}
\end{tabbing}
\noindent where {\it V}, {\it C}, {\it L} and {\it BV} are
addresses to which control must be transferred in case the
head normalized and dereferenced version of the value stored in
register {\it A1} is, respectively, a flexible term, a rigid term that
has a constant different from {\it ::} as its head, a nonempty list or
a term with a bound variable as its head. In the use that is made of
this instruction above, {\it fail} is assumed to be the location of
code that causes backtracking. The instructions {\it try\_me\_else}
and {\it trust\_me} that are used here function as they do in the WAM
to create, utilize and discard choice points; an extra numeric
argument has been included with each of them that indicates the number
of argument registers that are to be saved or retrieved as
relevant. The {\it unify\_variable} and {\it unify\_value}
instructions that are used take type parameters for reasons that we
have already explained. In this particular instance, {\it ty1} and
{\it ty2} are to be understood as references to the representation of
the types {\it i} and {\it (list i)}, respectively. We note that in
the only place where the {\it unify\_value} instruction appears in
this code, there is no utility for the type argument and, observing that
this instruction will never be executed in read mode, we may replace
it with a special {\it set\_value} instruction as suggested in
\cite{HAK91}.\footnote{This is, in fact, what is done in the abstract
machine and compilation model actually underlying the {\it Teyjus}
implementation.} As a final comment, we observe that both the {\it
proceed\_finish\_unify} and the {\it execute\_finish\_unify}
instructions that appear in this code are essential: depending on the
form of the first and third incoming arguments, execution of the term
simplification code for either clause may lead to bindings that affect
the state of the existing disagreement set.

The definition of ${\it mappred}$ presented in Section~\ref{sec:hohc}
illustrates a mixed use of a predicate variable. Compilation of that
definition might produce the following code: 
\begin{tabbing}
\qquad\={\it mappred:}\quad\={\it switch\_on\_term L2, L3, L4, fail}\quad\=\kill
\>{\it mappred:}\>{\it switch\_on\_term L2, L3, L5, fail}\>{\it \%}\\
\>{\it L2:}\>{\it try\_me\_else L4, 3}\>{\it \% mappred}\\
\>{\it L3:}\>{\it get\_nil A1}\>{\it \% nil}\\
\>\>{\it get\_nil A3}\>{\it \% P nil}\\
\>\>{\it proceed\_finish\_unify}\>{\it \%}\\
\>{\it L4:}\>{\it trust\_me 3}\>{\it \% mappred}\\
\>{\it L5:}\>{\it allocate}\>{\it \%}\\                     
\>\>{\it get\_list A1}\>{\% (:: }\\
\>\>{\it unify\_variable A4, ty1}\>{\it \% X}\\
\>\>{\it unify\_variable Y2, ty2}\>{\it \% L1)}\\
\>\>{\it get\_variable Y1, A2}\>{\it \% P}\\
\>\>{\it get\_list A3}\>{\it \% (:: }\\
\>\>{\it unify\_variable A2, ty1}\>{\it \% Y}\\
\>\>{\it unify\_variable Y3, ty2}\>{\it \% L2)}\\
\>\>{\it call\_finish\_unify 3, 3}\>{\it \% :- }\\
\>\>{\it globalize Y1, A3}\>{\it \%}\\
\>\>{\it put\_capp A1, A3, 2}\>{\it \% S1 = (P }\\
\>\>{\it unify\_value A4, ty1}\>{\it \% X}\\
\>\>{\it unify\_value A2, ty1}\>{\it \% Y)}\\
\>\>{\it call solve, 3}\>{\it \% S1, }\\
\>\>{\it put\_value Y2, A1}\>{\it \% (mappred L1}\\
\>\>{\it put\_value Y1, A2}\>{\it \% P}\\
\>\>{\it put\_value Y3, A3}\>{\it \% L2}\\
\>\>{\it deallocate}\>{\it \%}\\
\>\>{\it execute mappred}\>{\it \% )}
\end{tabbing}
\noindent We assume here that {\it ty1} and {\it ty2} are references
to the representation of the types {\it i} and {\it (list i)},
respectively. The flexible goal {\it (P X Y)} is translated in this
code by a call to the predicate {\it solve} as discussed earlier in
this section. Towards understanding the nature of 
this translation, we might consider the execution of the query
\begin{tabbing}
\qquad\=\kill
\> ${\it mappred} \app ({\it bob} :: {\it sue} :: {\it nil}) \app (\lambdax {x} \lambdax {y}
\somex {z} ({\it parent} \app x \app z) \land ({\it parent} \app z \app y))\app L$.
\end{tabbing}
\noindent discussed in Section~\ref{sec:hohc}. Clause indexing will
lead to the selection of the code for the second clause for {\it
mappred} in this case. The term simplification part of this code will
execute successfully, the term $\somex {z} ({\it parent} \app
{\it bob} \app z) \land ({\it parent} \app z \app y))$ will be formed
and stored in register {\it A1} and the code for {\it solve} will be
invoked. Using the definition of {\it solve}, this goal will be
simplified,  leading eventually to the invocation of the atomic goals
{\it (parent bob Z)} and {\it (parent Z Y)}. The recursive call to
{\it mappred} will lead, in a similar fashion, to the invocation of the
atomic goals {\it (parent sue Z')} and {\it (parent Z' Y')}. The query
variable {\it L} will be bound at the end to a list 
containing the values determined for {\it Z} and {\it Z'} by these
goals. Another point to note is that all the unification problems
that arise relative to the query of interest are ones that can be
solved without the invocation of the interpretive, higher-order
phase.

\section{Conclusion}\label{sec:conc}

We have considered in this paper the implementation of an extension to
logic programming that is based on permitting a quantification over
predicate and function symbols and on using lambda terms as data
structures in place of first-order terms. In addition to a careful
exposition of the issues that need to be dealt with in a low-level
realization of such an extension, our contributions are threefold: we
have discussed representations for lambda terms that facilitate their
intensional treatment, we have presented mechanisms for realizing term
reduction and for supporting higher-order unification within a logic
programming machine model and we have sketched an approach to
compilation. The ideas that we have presented here have been used in
amalgamation with other devices that we have developed for the
treatment of new scoping mechanisms and of polymorphic typing in an
actual implementation of the $\lambda$Prolog language.

A question often of interest in the context of language
enrichments is the performance degradation that is to be incurred on
account of them. There are two factors that lead to a different
treatment of first-order programs within our framework from that in
traditional Prolog implementations. First, as discussed in
Section~\ref{ssec:intrep}, a representation must be used for
compound terms that permits changes to be made to internal nodes in
their tree-like structure. Second, the occurs-check that is usually
omitted in logic programming languages is not really a luxury in the
important higher-order applications. A third factor, not discussed
here but that is relevant to the full $\lambda$Prolog language, is a
runtime overhead arising from polymorphic typing. 
The impact of the occurs-check is obviously non-uniform and therefore
impossible to quantify in a general manner. A careful assessment of the
first and third factors requires experiments with controlled auxiliary
implementations, something beyond the scope of this
paper. However, a rough assessment is possible. Lists receive a
specialized treatment in 
the {\it Teyjus} system that comes close to the usual representation
of first-order structures. By contrasting performance under such a
treatment with that when a vanilla functor-based representation is
used, a sense of the additional cost can be obtained.

\begin{table}
\caption{Timing Comparisons with Prolog Implementations over Naive Reverse}
\label{fig:nrev}
\begin{minipage}{\textwidth}
\begin{tabular}{lccc}
\hline\hline
{\it System}& {\it Special List} & \multicolumn{2}{c}{\it Functor Based}\\
{\it Employed} & {\it Representation} & \multicolumn{2}{c}{\it
Representation}\\
\hline
{\it Teyjus (v 1.0-b32)} & 11.99 secs & 18.67 secs &  21.18 secs\\
& {\it (Polymorphic)} &{\it\qquad (Monomorphic)} & {\it (Polymorphic)}\qquad\\
\noalign{\vspace {0.5cm}}
{\it SWI-Prolog (v 4.0.0)} & 8.1 secs & \multicolumn{2}{c}{8.8
secs}\\
\noalign{\vspace {0.5cm}}
{\it SICStus (v 3.9.1)} & 0.23 secs & \multicolumn{2}{c}{0.35 secs} \\
\hline\hline
\end{tabular}
\end{minipage}
\end{table}

Table~\ref{fig:nrev} presents the results of the kind of experiment
described above, performed with {\it Teyjus} version 1.0-b32 modified
to omit the occurs-check. The numbers in the table represent the time
taken by 10,000 invocations of naive reverse on a 30 element
list. All trials, here and below, were carried out on a 440 MHz
UltraSPARC-IIi processor. A functor-based representation for lists in
{\it Teyjus} can be chosen to be either monomorphic or polymorphic in
nature and execution times are provided for both.  In contrast, the
specialized list representation is available only in polymorphic
form. From the timing measurements for the polymorphic versions, we
conclude that there is about a 75\% overhead to not using the
specialized representation. This is appropriately viewed as an upper
bound on the additional cost for a higher-order representation, at
least some of the improved performance being attributable
to specialized compilation for lists. Polymorphism adds about a 13.5\%
overhead in the functor-based representation and we estimate a similar
cost under the special treatment of lists. For comparison, we also
present performance measurements for two Prolog implementations;
from the perspective of running time, these figures are
best thought of as applying to monomorphic list representations. The
contrast with {\it SICStus} is humbling, indicating the distance to go in
building a well-engineered and highly optimized implementation, even
if revealing little by way of the difference between treatments of
the first-order and the higher-order language.

Another important aspect of comparison is that of contrasting our
ideas and system with those of other implementations of
$\lambda$Prolog. There have been four previous implementations of this
language. Three of these are interpreter based, built using
Prolog~\cite{lp2.7}, Lisp~\cite{elliott89elp} and Standard
ML~\cite{EP91,WM97terzo}. None of these systems considered in any
detail the special issues that arise in a low-level treatment of the
higher-order aspects of $\lambda$Prolog and a comparison with them
therefore appears not to be very meaningful.\footnote{The performance
comparisons made in \cite{BR92b} with the Lisp version substantiate
this viewpoint.}  The only remaining realization of $\lambda$Prolog,
called {\it Prolog/Mali} \cite{BR92b}, is one that translates
$\lambda$Prolog programs into C code that can then be compiled. The
translation process utilizes a memory management system called {\it
Mali} that has been developed especially for logic programming
languages: in particular, translation is realized in the form of calls
to functions supported by this system. Using this approach has the
distinct benefit that a memory management scheme is automatically
available but it also forces some awkward choices, such as the full
copying of clause bodies, to be in consonance with the framework
provided by {\it Mali}.

Despite the difference in overall structure, there is a scheme to the
treatment of the higher-order aspects in {\it Prolog/Mali} that can be
compared with the ideas we have presented in this paper. At the level of
term representation, there seem to be three differences. First, the de
Bruijn scheme for rendering bound variables is rejected in {\it
Prolog/Mali} on the grounds that ``it forces to renumber the rightmost
term.'' While this observation is correct in principle, it appears not
to be relevant in practice as we have pointed out in
Section~\ref{ssec:debruijn}. To support the comparison of terms in a
situation where a name-based encoding is used for bound variables, an
approach based on using new constants is suggested. Unfortunately, the 
details of this approach are not explained completely making a
satisfactory assessment of it impossible.\footnote{There are also vestiges
of this approach in answer presentations that remain unclear to us
and, quite possibly, to other $\lambda$Prolog users.} A second difference
is that an explicit substitution mechanism is not considered in
{\it Prolog/Mali} and reduction substitutions seem to be effected
eagerly. Finally, first-order terms seem to obtain the usual
Prolog-like treatment in {\it Prolog/Mali}, higher-order facets being
handled via special attributes attached to terms. The treatment of
higher-order 
unification and the integration of reduction into the overall
computational model receives little discussion in \cite{BR92b} and, in
light of this, we believe that a detailed consideration of these
aspects is unique to our work; an interesting exception, however, is
the idea of indexing flexible-flexible pairs by their flexible heads,
to be awakened by bindings for these heads, a possibility whose
integration into our processing model bears investigation. The last
relevant aspect is the compilation of unification. Clearly, the
underlying machine model is explicitly manifest only in our work
although many ideas relating to the compilation of the first phase of
simplification of disagreement sets receive a similar treatment in
both contexts and share also with an early presentation of some of
our ideas \cite{NJ89}. 

\begin{table}
\caption{Timing Comparisons between Teyjus and Prolog/Mali}
\label{fig:pmcompare}
\begin{minipage}{\textwidth}
\begin{tabular}{lcc}
\hline\hline
{\it System} & {\it Naive Reverse} & {\it Type Inference}\\
\hline
{\it Teyjus (v 1.0-b32)} & 11.99 secs & 2.95 secs \\
\noalign{\vspace{0.5cm}}
{\it Prolog/Mali } & 12.00 secs & 9.59 secs\\
\hline\hline
\end{tabular}
\end{minipage}
\end{table}

Table~\ref{fig:pmcompare} complements our qualitative comparisons by
presenting execution times for {\it Prolog/Mali} and {\it Teyjus} on two
different kinds of tasks. The {\it naive reverse} program is the one
used in the earlier tests and, as such, provides a measure of
behaviour over first-order programs. The {\it type inference} program
assigns type schemes to ML-like programs and is a good example for
testing performance over higher-order terms, reduction and (a
specialized form of) higher-order unification.\footnote{This program
also includes dynamically scoped constants and assumptions and
performance differences are therefore not entirely attributable to
the treatment of features discussed in this paper. However, we 
intend the figures that we present to be suggestive rather than
definitive for the reasons we explain.} 
The indications from these tests is that the {\it Teyjus} system
matches performance of {\it Prolog/Mali} over first-order programs and
does significantly better on genuine higher-order ones. A larger set
of tests is needed 
to draw more substantive conclusions. Unfortunately, there are
practical difficulties to providing a suitable collection that
highlights genuine performance differences. {\it Prolog/Mali} omits
the occurs-check that is significant to higher-order applications, 
uses a non-standard syntax for $\lambda$Prolog programs leading to a
substantial overhead in adapting available user programs to run
under it and, finally, appears to yield incorrect results in a few
of the examples we tried.

Our focus in this paper has been on describing a broad framework for
the treatment of higher-order features in logic programming. There are
obviously tradeoffs in the actual deployment of these ideas. Although
beyond the scope of the present study, a quantification of these
tradeoffs is important and is, in fact, the object of other work
\cite{LN02,LNX03,NX03}. A particularly exciting  
direction that we are now exploring is that of fine-tuning our
abstract machine and compilation model to the important subclass of
higher-order programs referred to as L$_\lambda$ programs
\cite{miller91jlc}, possibly even with some loss of completeness over
the full collection. In a different vein, many of our
implementation ideas are applicable in related contexts, such as that
of logic programming within a dependently typed lambda calculus
\cite{Pfenning99cade}. The extension of this work in these directions
is also a matter under investigation. 

\section*{Acknowledgements}

Earlier versions of the ideas in this paper have benefitted from
discussions with Bharat Jayaraman and Debra Wilson. Comments from the
reviewers have helped in an improved presentation.

\end{document}